\documentclass{fac}

\usepackage{xy}
\xyoption{all}

\usepackage{amssymb}
\usepackage{amstext}
\usepackage{amsmath}
\usepackage{stmaryrd}

\usepackage{color}
\usepackage{pstricks}
\usepackage{rotating}

\usepackage{multirow}

\newcommand{\Chan}{\CCC}
\newcommand{\Chty}{\Theta}
\newcommand{\Conf}{\PPP}
\newcommand{\Nod}{\NNN}
\newcommand{\Ide}{\JJJ}
\newcommand{\contr}{\copyright}
\newcommand{\contrr}{\circledR}
\newcommand{\Human}{I}
\newcommand{\Device}{D}

\newcommand{\then}{\Longrightarrow}
\newcommand{\rar}{\shortrightarrow}

\newcommand{\orig}[1]{\sqrt{{#1}}}
\newcommand{\Knows}{\Gamma}
\newcommand{\probe}{sample}
\newcommand{\prob}{sample}
\newcommand{\appear}{emit}

\newcommand{\sample}{source}

\newcommand{\Local}{{\sf Local}}
\newcommand{\Global}{{\sf Global}}

\newcommand{\WP}{\mbox{\Large $\wp$}}

\newcommand{\Var}{\XXX}

\newcommand{\tto}[1]{\xrightarrow{#1}}

\newcommand{\oot}[1]{\xleftarrow{#1}}

\newcommand{\flow}[1]{\stackrel{#1}\longrightarrow}

\newcommand{\writ}[1]{\left<\, #1\,\right>}
\newcommand{\wwrit}[1]{\left<\left<\, #1\,\right>\right>}
\newcommand{\reaa}[1]{\left(\, #1\,\right)}
\newcommand{\send}[1]{\left<\cdot\, #1\, \cdot \right>}
\newcommand{\trm}[1]{\left[\, #1\, \right]}
\newcommand{\recv}[1]{\left(\cdot\, #1\, \cdot \right)}
\newcommand{\prb}[1]{\left(: #1:\right)}
\newcommand{\apr}[1]{\left<: #1:\right>}
\newcommand{\ssend}[1]{\left<\left<\cdot\ #1\ \cdot \right>\right>}
\newcommand{\rrecv}[1]{\left(\left(\cdot\ #1\ \cdot \right)\right)}
\newcommand{\pprb}[1]{\left(\left(: #1:\right)\right)}
\newcommand{\aapr}[1]{\left<\left<: #1:\right>\right>}

\newcommand{\tagg}[1]{\tag{\sf{#1}}}

\newcommand{\CCC}{{\cal C}}

\newcommand{\EEE}{{\cal E}}
\newcommand{\FFF}{{\cal F}}

\newcommand{\JJJ}{{\cal J}}

\newcommand{\LLL}{{\cal L}}

\newcommand{\NNN}{{\cal N}}

\newcommand{\PPP}{{\cal P}}

\newcommand{\TTT}{{\cal T}}

\newcommand{\XXX}{{\cal X}}

\renewcommand{\Bbb}{\mathbb}

\newcommand{\EEe}{{\Bbb E}}
\newcommand{\FFf}{{\Bbb F}}

\newcommand{\TTt}{{\Bbb T}}

\mathcode`\<="4268 
\mathcode`\>="5269 
\mathchardef\gt="313E 
\mathchardef\lt="313C 

 \def\pushright#1{{
    \parfillskip=0pt            
    \widowpenalty=10000         
    \displaywidowpenalty=10000  
    \finalhyphendemerits=0      
    \leavevmode                 
    \unskip                     
    \nobreak                    
    \hfil                       
    \penalty50                  
    \hskip.2em                  
    \null                       
    \hfill                      
    {#1}                        
    \par}}                      

 \def\qed{\pushright{$\square$}\penalty-700 \smallskip}

\newenvironment{prf}[1]{\begin{trivlist} \item[{\bf ~Proof}#1.]}%
{\qed\end{trivlist}}

\newcommand{\be}[1]{\begin{#1}}

\newcommand{\ee}[1]{\end{#1}}
\newcommand{\beq}{\begin{equation}}
\newcommand{\eeq}{\end{equation}}
\newcommand{\ba}[1]{\begin{array}{#1}}
\newcommand{\ea}{\end{array}}
\newcommand{\bea}{\begin{eqnarray}}
\newcommand{\eea}{\end{eqnarray}}
\newcommand{\bear}{\begin{eqnarray*}}
\newcommand{\eear}{\end{eqnarray*}}
\newcommand{\bpr}{\begin{prf}{}}
\newcommand{\epr}{\end{prf}}
\newcommand{\bprf}[1]{\begin{prf}{#1}}
\newcommand{\eprf}{\end{prf}}

\newtheorem{thm}{Theorem}[section]
\newtheorem{defn}[thm]{Definition}

\newtheorem{cond}{}[thm]

\newtheorem{prenumb}[thm]{\hspace{-1ex}}

\newcounter{countroman}
{\begin{list}{{\rm (\roman{countroman})}}{\usecounter{countroman}}}%
{\end{list}}

\newcounter{countalpha}
\newenvironment{anumerate}%
{\begin{list}{(\alph{countalpha})}{\usecounter{countalpha}}}%
{\end{list}}

\newcounter{countalphabf}
{\protect\begin{list}{{\rm (}{\bf \protect\alph{countalphabf}}{\rm%
)}}{\protect\usecounter{countalphabf}}}%
{\end{list}}

\title{Actor-network procedures: \\
Modeling  multi-factor authentication, device pairing, social interactions}

\author[D. Pavlovic and C. Meadows]{Dusko Pavlovic$^1$ 
and  Catherine Meadows$^2$ \\
$^1$Royal Holloway University of London and Universiteit Twente; Email: {\tt 
dusko.pavlovic\char64 rhul.ac.uk}
 \\
$^2$Naval Research Laboratory, Washington, DC, USA; 
Email: 
{\tt meadows@itd.nrl.navy.mil}
}

\begin{document}

\maketitle

\begin{abstract}
As computation spreads from computers to networks of computers, and migrates into cyberspace, it ceases to be globally programmable, but it remains programmable indirectly and partially: network computations cannot be controlled, but they can be steered by imposing local constraints on network nodes. The tasks of "programming"  global behaviors through local constraints belong to the area of \emph{security}. The ``program particles" that assure that a system of local interactions leads towards some desired global goals are called \emph{security protocols}. They are the software connectors of modern, world wide software systems.

As computation spreads beyond cyberspace, into physical and social spaces, new security tasks and problems arise. As computer networks are extended by nodes with physical sensors and controllers, including the humans, and interlaced with social networks, the engineering concepts and techniques of computer security blend with the social processes of security, that evolved since the dawn of mankind. These new connectors for computational and social software require a new ``discipline of programming" of global behaviors through local constraints. Since the new discipline seems to be emerging from a combination of established models of security protocols with older methods of procedural programming, we use the name \emph{procedures} for these new connectors, that generalize protocols.

In the present paper we propose \emph{actor-networks} as a formal model of computation in heterogenous networks of computers, humans and their devices, where these new procedures run; and we introduce \emph{Procedure Derivation Logic} (PDL) as a framework for reasoning about security in actor-networks. On the way, we survey the guiding ideas of \emph{Protocol Derivation Logic} (also PDL) that evolved through our work in security in last 10 years. Both formalisms are geared towards graphic reasoning and, ultimately, tool support. We illustrate their workings by analysing a popular form of two-factor authentication, and a multi-channel device pairing procedure, devised for this occasion.
\end{abstract}

%
%
%

\section{Introduction}
\subsection{Motivation and background}
In \cite{PavlovicD:SEFM10} we pondered about the ``unreasonable ineffectiveness of security engineering", and suggested that one of the main causes was that the widely used methods for pervasive software design were low level. The high level methodologies to specify and design reusable software procedures were not lifted from traditional computer systems to modern network based computation. In Section~IV of \cite{PavlovicD:SEFM10}, we provided a sketch of a network computation model that might fill the gap, and be used as a high level tool to specify and analyze network procedures. But the sketch was very crude, and we did not even have space to provide any examples of network procedures. So the final part of that story remained rather obscure. We attempt to rectify that in the present paper.

\subsection{Context}
\subsubsection{The ages of software}
In the beginning, engineers built computers, and wrote 
programs to control computations. The platform of computation was the computer,  and it was used to execute algorithms and calculations, allowing people to discover, e.g., fractals, and to invent  compilers, that allowed them to write and execute more 
algorithms and more calculations more efficiently. Then the 
operating system became the platform of computation, and 
software was developed on top of it. The era of personal 
computing and enterprise software broke out. And then the 
Internet happened, followed by cellular networks, and 
wireless networks, and ad hoc networks, and mixed networks. Cyber space emerged as the distance-free space of instant, costless communication, where any pair of network nodes is directly connected. 
Nowadays software is developed to run in cyberspace. The Web is, strictly speaking, just a software system, albeit a formidable one. A botnet is also a software system. As social space blends with cyber space, many social (business, collaborative) processes can be usefully construed as software systems, that ran on social networks as hardware. Many social and computational processes become inextricable. Table~\ref{ages-comp} gives a crude picture of the paradigm shifts that led to this remarkable situation.

\begin{table}
\begin{center}
\begin{oldtabular}{|c||c|c|c|c|}
\hline
\textit{\textbf{age}} & \textit{ancient times} & \textit{middle ages} & \textit{modern times} \\[1ex]
\hline \hline
\textbf{platform}  & computer & operating system & network \\[1ex]
\hline
 \textbf{applications} & Quicksort, compilers & MS Word, Oracle  & 
WWW, botnets  \\[1ex]
\hline
\textbf{requirements} & correctness, termination & liveness, safety 
& integrity, confidentiality \\[1ex]
 \hline
\textbf{tools} & programming languages & specification languages & scripting 
languages \\[1ex]
 \hline
\end{oldtabular}
\end{center}
\caption{Paradigms of computation}
\label{ages-comp}
\end{table}%

But as every person got connected to a computer, and every computer to a  network, and every network to a network of networks,  computation became interlaced with communication, and ceased to be programmable.  The functioning of the Web and of web applications is not determined by the code in the same sense as in a traditional software system: after all, web applications do include the human users as a part of their runtime. The fusion of social and computational processes in cyber-social space leads to a new type of information processing, where the purposeful program executions at the network nodes are supplemented by spontaneous data-driven evolution of network links. While the network emerges as the new computer, data and metadata become inseparable, and new types of security problems arise.

\subsubsection{The ages of software security}
In early computer systems, security tasks mainly concerned sharing of the computing resources. In computer networks, security goals expanded to include information protection. Both computer security and information security essentially depend on a clear  distinction between the secure areas, and the insecure areas, separated by a security perimeter. Security engineering caters for computer security and for information security by providing the tools to build the security perimeter. In cyber space, the secure areas are separated from the insecure areas by the ``walls" of cryptography; and they are connected by the ``gates" of cryptographic protocols.  

But as networks of computers and devices spread through physical and social spaces, the distinctions between the secure and the insecure areas become blurred.  With network computation, the software-hardware distinction acquires a new meaning. In contrast with the purposefully built and programmed electronic computers, the new spontaneously evolving computer-as-a-network includes social networks as  a part of its hardware, while social processes are becoming a part of its software. To follow these developments, computer science is endorsing themes and tools of social sciences \cite{WattsD:book,KleinbergJ:nets-book}, while social sciences are increasingly concerned with computation \cite{WellmanB:compnets-socnets,BenklerY:book}. The formalism of \emph{actor-networks} arises on this background.

\begin{table}
\begin{center}
\begin{oldtabular}{|c||c|c|c|c|}
\hline
\textit{\textbf{age}} &  \textit{middle ages} & \textit{modern times} & \textit{postmodern times} \\[1ex]
\hline \hline
\textbf{space}  & computer center & cyber space & cyber-social space \\[1ex]
\hline
 \textbf{assets} & computing resources  & 
information & public and private resources  \\[1ex]
\hline
\textbf{requirements} & availability, authorization & integrity, confidentiality 
& trust, privacy \\[1ex]
 \hline
\textbf{tools} &  locks, tokens, passwords & cryptography, protocols & mining and classification \\[1ex]
 \hline
\end{oldtabular}
\end{center}
\caption{Paradigms of security}
\label{ages-sec}
\end{table}%

\subsection{Goals and ideas}\label{Goals}
Our goal is to contribute towards a formal framework for  for \emph{reliable} and \emph{practical} reasoning about computation and communication in networks. Since network computation involves adversarial behaviors, security stands out as the central concern in network computation. But reliable reasoning about security, even in the familiar end-to-end networks, usually requires complicated models. Hence the tension between the requirements of reliability and  of practicality: reliable reasoning requires a precise formal model, but formal models of security tend to be impractically complex. One approach is to mitigate this complexity through automated support. We have been studying this solution for many years \cite{MeadowsC:CSFW91,Meadows96,Meadows07,PavlovicD:ARSPA06}, and it has been broadly supported in the research community \cite{Abadi-Blanchet-Lundh,Bella:book,Basin:decade,Cortier-Kremer:book}. Another approach is to try to decrease the complexity through model abstraction and refinement, and through search for convenient and intuitive notations. In the present work, we put more emphasis on this second approach, building upon our previous  attempts \cite{PavlovicD:ESORICS04,PavlovicD:CSFW05,PavlovicD:ESORICS06} to extend to reasoning about security the incremental modeling methodologies, well established in software engineering. Towards this goal, we draw our formal models from the informal reasoning practices, and attempt to make them mathematically precise, while trying to keep them as succinct and intuitive as possible. The main feature of our formalism is that it provides support for \emph{diagrammatically} based security proofs. Although they cannot be directly automated, these proofs are as formal and as precise as the proofs in similar diagrammatic formalisms across mathematics, which also cannot be directly automated. After all, very few of the formal proofs presented in mathematics papers and textbooks are ``formal enough" to be entered into a theorem prover.  The hope is, however, that in the end, the two types of security models, those designed for software tools, and those designed for human consumption, will converge into a theory that will allow automating complex arguments, while resolving some complexities through insightful notations.

An important instrument of user-friendly mathematical formalisms is the ``syntactic sugar", where we subsume a whole gamut of notational and graphical abbreviations, conventions and abuses. Although unsound informal reasoning can be a source of many troubles, and the pedagogical emphasis is usually placed squarely against it, sound informal reasoning can be a source of many insights. After all, the formal proofs and constructions are seldom born fully shaped and formalized, but begin their life as insights and ideas. A useful formalism supports such transformations from insights into formal proofs. The soundness of such transformations often depends on a natural selection of notational conventions and abuses. Not entirely unintentionally, our formalisms turn out to be rich in graphic and syntactic sugar, and in sound notational abuses. We begin by explaining some terminological abuses, starting from the first three words of the title. 

\subsubsection{``Actor-network"}
Networks have become an immensely popular model of computation across sciences, from physics and biology, to sociology and computer science \cite{Dorogovtsev-Mendes,networks-biology,NewmanM:book}. Actor-networks \cite{Latour:reassembling} are a particularly influential paradigm in sociology, emphasizing and analyzing the ways in which the interactions between people and objects, as equal factors, drive social processes, in the sense that most people cannot fly without an airplane; but that most airplanes also cannot fly without people. Our goal in the present paper is to formalize and analyze some security processes in networks of people, computers, and the ever expanding range of devices and objects used for communication and networking, blurring many boundaries. The idea that people, computers, and objects are equal actors in such networks imposed itself on us, through the need for a usable formal model, even before we had heard of the sociological actor-network theory. After we heard of it, we took the liberty of adopting the name actor-network for a crucial component of our mathematical model, since it conveniently captures many relevant ideas. While the originators of actor-network theory never proposed a formal model, we believe that the tasks, methods and logics that we propose are not alien to the spirit of their theory. In fact, we contend that computation and society have pervaded each other to the point where computer science and social sciences already share their subject. 

It should be noted, though, that the goals of this work are completely different from the goals of sociology of actor-networks, and that our actor-network formalism deviates from the original ideas in a substantial way, even by being a formalism. We make no claims or attempts to faithfully interpret any of the actor-network authors; but we remain faithful to the spirit of their endeavor, since they all discourage orthodoxy.

\subsubsection{``Procedures"}
In computer programs, frequently used sequences of operations are encapsulated into \emph{procedures}, also called \emph{routines}. A procedure can be called from any point in the program, and thus supports reuse of code.

In computer networks, frequently used sequences of operations are specified and implemented as \emph{network protocols}, or as \emph{cryptographic protocols}. So protocols are, in a sense, network procedures. Conceptually, if not technically, protocol analysis can thus be viewed as an extension of the venerable science of program semantics, and of the methods of procedural programming, adapted for the purposes of network computation.

Beyond computer networks, there are now hybrid networks, where besides computers with their end-to-end links, there may be diverse devices, with their heterogenous communication channels, cellular, short range etc. Online banking and other services are nowadays usually secured by two-factor and multi-factor authentication, combining passwords with smart cards, or cell phones. A vast area of \emph{multi-channel} and \emph{out-of-band} protocols opens up, together with the web service \emph{choreographies} and \emph{orchestrations}; and we have only scratched its surface. And then there are of course also social networks, where people congregate with their phones, their cameras and their smiling faces, and overlay the wide spectrum of their social channels over computer networks and hybrid networks. Many sequences of frequently used operations within these mixed communication structures have evolved. This is what we call \emph{actor-network procedures}. 

Conceptually, actor-network procedures extend program procedures from computers to networks; and they furthermore extend network protocols, and multi-channel protocols, and web service choreographies and orchestrations, into social networks. 

Technically, actor-network procedures are, of course, immensely more complicated than their conceptual relatives in computer science, because humans use many types of physical resources, communicate through many parallel communication channels, with many levels of encoding interleaved over each other. If we ignore computers and devices, actor-network procedures already capture the main complexities of social life, as actor-network theorists are explaining. They are a sociologists' problem. In any case, they are completely out of reach for computer scientists' protocol models, symbolic, information theoretic, and computational, all designed for simple end-to-end communication. So a computationally minded reader may wonder why bother to bring them up here.

The reason is that the most frequent transactions that we engage with our computers and even among ourselves involve actor-network procedures. Online banking and shopping, as well as the checkout in the supermarket, travel search on the web, as well as the security line on the airport involve actor-network procedures. Time and again, it has been recognized and reconfirmed that most security breaches nowadays occur not through cryptanalysis, and not through buffer overflow, but through various forms of ``social engineering" and channel interactions. Yet ``social engineering" has remained a marginal note in technical research. An effort to change this may be foolhardy, but it leads to ideas and structures that seem interesting to explore --- even independently of their utility.

\subsection{Related work}
\subsubsection{The expanding concept of a protocol}
In social and computational networks, procedures come in many flavors, and have been studied from many angles. Besides cryptographic protocols, used to secure end-to-end networks, in hybrid networks we increasingly rely on multi-channel protocols \cite{Stajano}, including device pairing \cite{Saxena:survey}. In web services, standard procedures come in two flavors: choreographies and orchestrations \cite{PeltzC:orchestrations}. There are, of course, also social protocols and social procedures, which were developed and studied first, although not formally modeled. As social networks are increasingly supported by electronic networks, and on the Web, social protocols and cryptographic protocols often blend together. Some researchers have suggested that the notion of protocol should be extended to study such combinations \cite{Blaze,EllisonC:ceremonies,TygarJ:ceremonies}. On the other side, the advent of ubiquitous computing has led to extensive, careful, but largely informal analyses of the problems of device pairing, and of security interactions of using multiple channel types \cite{Stajano,Hoepman05,Roscoe}.  One family of the device pairing proposals has been systematically analyzed in the computational model in \cite{Vaudenay:SAS,Passini06:SAS,Laur-Nyberg,Laur-Passini}.

\subsubsection{Protocol logics and graphics}
\label{PCL-PDL}

There is a substantial and extremely successful body of research on the formal specification and verification of security protocols. As we have remarked, it is largely geared to supporting
sound and efficient mechanisms for specification and verification, while considerably less attention has been paid to approaches that support the user's understanding of the
structure of a protocol and how it contributes to its security.  There have been some notable exceptions, however.  In this section we describe the work in this
direction that has contributed to our own efforts.

One of the most successful, and in our opinion most interesting formal methods for reasoning about security protocols are \emph{strand spaces} \cite{strands}. Among its many salient features, the convenient diagrammatic protocol descriptions were an important reason for its wide acceptance and popularity. It is important to note that the strand space diagrams are not just an intuitive illustration, but that they are formal objects, corresponding to precisely defined components of the theory, while on the other hand closely resembling the informal ``arrows-and-messages" protocol depictions, found in almost every research paper and on almost every white board where a protocol is discussed.

Protocol Composition Logic (PCL) was, at least in its early versions \cite{PavlovicD:CSFW01,PavlovicD:CSFW03,PavlovicD:MFPS03,PavlovicD:JCS04,PavlovicD:JCS05}, an attempt to enrich the strand model with a variable binding and scoping mechanism, making it into a process calculus with a formal handle on data flows, which would thus allow attaching Floyd-Hoare-style annotations to protocol executions, along the lines of \cite{PavlovicD:ASE01,PavlovicD:AMAST02}. This was necessary for incremental refinement of protocol specifications, and for truly compositional, and thus scalable protocol analyses, which were the ultimate goal of the project. Unfortunately, with these extensions, the handy diagrammatic notation of the strand model got lost.

Protocol Derivation Logic (PDL) has been an ongoing effort  \cite{PavlovicD:ESORICS04,PavlovicD:CSFW05,PavlovicD:ESORICS06,PavlovicD:ARSPA06,PavlovicD:dist06,PavlovicD:MFPS10} towards a scalable, i.e. incremental protocol formalism, allowing composition and refinement like PCL, but equipped with an intuitive and succinct diagrammatic notation, like strand spaces. The belief that these two requirements can be reconciled is based on the observation that the reasoning of protocol participants is concerned mostly with the order of events in protocol executions.\footnote{E.g., in order to achieve mutual authentication, each participant of a run must be able to prove that their and their peers' actions in their conversation must have happened exactly in the order prescribed by the protocol: i.e., that the received messages were previously sent as claimed, and vice versa.} It follows that the protocol executions and their logical annotations both actually describe the same structures, which can be viewed as partially ordered multisets \cite{PrattV:Pomsets}, and manipulated within the same diagrammatic language. This has been the guiding idea of PDL. Several case studies of standard protocols, and the taxonomies of the corresponding protocol suites, have been presented in \cite{PavlovicD:ESORICS04,PavlovicD:CSFW05,PavlovicD:ESORICS06,PavlovicD:ARSPA06}. An application to a family of distance bounding protocols has been presented in \cite{PavlovicD:dist06}; and an extension supporting the probabilistic reasoning necessary for another such family has been proposed in \cite{PavlovicD:MFPS10}. In the present paper, we propose the broadest view of PDL so far --- which should here be read as \emph{Procedure} Derivation Logic. The underlying process model is enriched by a new network model, to support reasoning about network procedures. The logical annotations are extended accordingly ---  still geared towards the diagrammatic reasoning, which still seems like a reasonable strategy, since principals' reasoning towards security remains largely concerned with order of actions.

\subsubsection{Computational soundness?}
Security is an old social process, but it is a relatively new technical problem. As many other new problems, it often looks unreasonably complicated: a couple of lines of a protocol can conceal a subtle problem for many years. This is sometimes mentioned as the characterizing feature of the field of security. A direct analysis leads to convoluted reasoning, often with an exponential explosion of the cases to be considered. As mentioned in Sec.~\ref{Goals}, this naturally leads to the idea of automated reasoning \cite{Meadows-Millen-Kemmerer}, which we pursued through several different frameworks \cite{MeadowsC:CSFW91,Meadows94,PavlovicD:ARSPA06,Meadows07}. As this idea came to be a widely accepted, it led to a convergence of research to a small number of standard formalisms, based on a sharp division between the symbolic and the computational models of security. This division originally started from the empiric observation, that motivated the seminal paper \cite{Abadi-Rogaway}, that the extant security research was broadly based on two different models. Since the computational model is more precise, whereas the symbolic model is easier to use, the way to get the best of both worlds is to demonstrate that  the proofs in the simpler model remain valid in the more precise model; in other worlds, that the symbolic model was \emph{computationally sound}. This approach led to many important results and useful tools \cite{Blanchet08,BartheG:CSF10,WarinschiB:survey}. But the focus on computational soundness led some researchers to begin viewing all formal models of security as approximations of the standard computational model, forgetting that all models are approximations of some real processes. In the ensuing confusion, even the models involving the features that ostensibly go beyond the computational model (such as timed channels \cite{PavlovicD:MFPS10}) were required to demonstrate their computational soundness. This is, of course, a meaningless requirement, since a model can only be sound or unsound with respect to a model where it can be faithfully interpreted.


Our current effort is again of this type, as the actor-network model includes several features that preclude computational interpretations, and render the question of its computational soundness meaningless. One such feature is the fact that the actors needn't be standard computers: in this paper, we will see networks involving humans, with their free will; but they could also be, e.g. ants, drawing unusual computational powers from their pheromones \cite{DorigoM:book}. Indeed, a framework that attempts to capture some social interactions, as announced in the title of this paper, can not be captured by a purely computational model, and thus can hardly be expected to be computationally sound. On the other hand, even if we accept to simulate all actors by Turing machines, including ants and humans, the resulting actor-network will still not boil down to the computational model, since the diverse communication channels, that can be specified in an actor-network, cannot be reduced to interactions of Turing machines. This is further discussed in Sec.~\ref{Actor-network model}. 

Undoubtedly, the fact that our model cannot be proven computationally sound, or even given a computational interpretation, can be interpreted as the evidence that we are modeling what cannot be modeled; whereas the fact that our proofs cannot be automated can be viewed as the evidence that they are not completely formal. Both interpretations are true, for some suitable meanings of the words ``formal", and ``model". Instead of arguing whether these meanings are reasonable or not, we present our formal model. We contend that our actor-network based proofs in Procedure Derivation Logic are as rigorous as the proofs in any of the standard mathematical formalisms; and hopefully somewhat insightful for the reader. In particular, our diagrammatic proofs can be viewed as a formal method similar to \emph{diagram chasing} in category theory \cite{MacLane,PavlovicD:MapsII}, from which we drew inspiration. 

\subsection*{Outline of the paper}
Sec.~\ref{Actor-network model} introduces the formal model of actor-networks. Sec~\ref{Processes-sec} explains how actor-networks compute, and introduces the formalisms to represent that computation, all the way to actor-network procedures. Sec.~\ref{Logic-sec} presents Procedure Derivation Logic (PDL) as a method for reasoning about actor-network procedures. In Sec.~\ref{Examples-sec} we provide the first case studies using PDL: we analyze the two-factor authentication in online banking, and a device pairing procedure combining physical and biometric channels. Sec.~\ref{Conclusions-sec} contains a discussion of the results and the future work.

\section{Actor-network model}\label{Actor-network model}
\subsection{A computer is a network is a computer}
The standard model of computation is a Turing machine. It uses a tape as a storage medium. Sometimes additional tapes are used to represent the input and the output interfaces. Probabilistic Turning machines also read some random  strings from a devoted tape, whereas oracle Turing machines communicate with the oracle through an additional tape. Last but not least, interactive computation is often modeled using several Turing machines that interact with each other on joint tapes. Such joint tapes are actually the \emph{communication channels} between the Turing machines.

Interactive Turing machines can be viewed as a computational network, with the machines as nodes, and the joint tapes as links between them. However, pervasive networks that compute in the world around us nowadays include a wide variety of computational agents at their nodes, including humans, and the various devices with different computational powers. Luckily, an abstract view of the nodes suffices for most analyses. We usually just need to specify that the state, i.e. the variables where the node can store its data; and we postulate which computations can be effectively performed by the node, and which computations are unfeasible.

The channels between the nodes can be viewed as a generalization of the joint tapes, shared by Turing machines in their interactions. But while the joint tapes trivially pass information from one machine to the other, nontrivial channels perform nontrivial data transformations. E.g., a pair of nodes (which may or may not be Turing machines) can be connected by a noisy binary channel, flipping each bit with a certain probability while passing it from one node to the other. Another pair of computational agents can be connected by a cyber channel, controlled by an adaptive attacker, who can change and modify the data flows at will. In symbolic protocol analysis, such attackers are usually specified as simple state machines. In cryptography, they are usually modeled as probabilistic polynomial-time Turing machines. In network models, it is convenient and intuitive to view them as processes encapsulated in channels.

In summary, the simplest model of network computation consists of
\begin{itemize}
\item computational agents, some of them controlled by various parties, others available as resources; and
\item communication channels between the agents, supporting different types of information flows.
\end{itemize}

\subsection{Idea of actor-networks}
Actor-networks depict social processes as computations, and computation as a social process. An example of an actor-network is a configuration consisting of a musician and her instrument. Their intended interaction is the music. The process and the result are highly structured, and the network representation helps with the analysis. The network first of all displays the symmetry of this interaction: the musician cannot play without the instrument, and the instrument cannot play without the musician. The fact that some musician's instrument may be a part of her body (e.g., her voice), and that some instrument's musician may be a computer makes the picture only more interesting. A smart card and a smart card reader form another network of this type. But this network is complicated by the need for someone to key in the pin of the card. The network grows still further if the card reader is connected to a bank, and perhaps dispenses money.

When networks involve heterogenous nodes, and heterogenous communication channels, then the diverse computational resources lead to different computational powers. This is where network computation essentially deviates from machine computation, where according to Church's Thesis, all the different machines have the same computational powers. In a network, one computer may be equipped with a camera and may provide a visual channel to a remote user, whereas another computer may be equipped with sensors and controllers, allowing it to stabilize the flight of an aircraft. A  smart card can perform some cryptographic operations when inserted in a reader, and other ones in the contactless mode. A musical instrument can produce one type of music with one musician, and something completely different with somebody else. The musician also depends on the instrument. Furthermore, after they are configured with their instruments, the musicians may further configure themselves into a higher-order configuration: an orchestra. And the orchestra can also be viewed as an actor within the configuration of an opera performance\ldots

Such configurations are what we call \emph{actor-networks}. A computational agent who participates in a configuration is an \emph{actor}, in the sense that she plays a particular \emph{role} assigned to it by a particular network \emph{procedure}. As computational networks spread and diversify, it is becoming increasingly important, and increasingly difficult, to assure that procedures provide the desired actor and network behaviors. Towards this goal, we formalize the above intuitions about actor-networks, and build a framework for reasoning about their procedures.

\paragraph{Remark.} The hierarchical structure of our actor-network formalism is alien to the spirit and the letter of original actor-network idea from sociology \cite{Latour:reassembling}. But it is essential for the goals of our logical analyses, which are different from the goals of sociological analyses. It may be of interest to explore the relation between the two sets of goals.

\subsection{Formalizing actor-networks}
\begin{defn}\label{AN-def}
 An \emph{actor-network} consists of the following sets:
\begin{itemize}
\item \emph{identities}, or \emph{principals} $\Ide = \{A,B,\ldots\}$,
\item \emph{nodes} $\Nod = \{M,N,\ldots\}$,
\item \emph{configurations} $\Conf = \{P,Q, \ldots\}$, where a configuration can be
\begin{itemize}
\item a finite set of nodes, or
\item a finite set of configurations;
\end{itemize}
\item \emph{channels} $\Chan = \{f,g,\ldots\}$, and
\item \emph{channel types} $\Chty = \{\tau,\varsigma,\ldots\}$
\end{itemize}
given with the following structure:
\[
\Chty \oot\vartheta \Chan \overset{\delta}{\underset{\varrho}{\rightrightarrows}} \Conf \stackrel{\contr} \rightharpoonup \Ide
\]
where
\begin{itemize}
\item the partial map $\contr:\Conf \rightharpoonup \Ide$ tells which principals control which configurations,
\item the pair of maps $\delta, \varrho : \Chan \to \Conf$ assign to each channel $f$ an \emph{entry} $\delta f$ and an \emph{exit} $\varrho f$, and
\item the map $\vartheta: \Chan \to \Chty$ assigs to each channel a type.
\end{itemize}

An \emph{actor} is an element of a configuration.
\end{defn}

\paragraph{Actors formally.} By the above definition, a configuration is thus   a tree whose leaves are annotated by network nodes. A configuration is an actor when viewed as an element of another configuration. In other words, an actor is formally a maximal subtree of a configuration tree. It is useful to distinguish the trees that come together to form another tree because this is what they do to perform some action together. For instance, a hand, a pencil and a piece of paper come together as actors to record a thought. A hand itself is a configuration of fingers, which come together to hold the pencil. The pencil is a configuration of graphite and wood. A door, a lock and a key come together to enforce someone's authority over a space. The lock is a configuration of its metal components. The writing configuration and the locking configuration come together to  put a novel in a drawer. And all of it is just trees.


\paragraph{Notation.} We denote by $N_B$ a node $N$ controlled by the principal $\contr N = B$. We write $g = (P \tto{\tau} N_B)$ for a channel $g$ of type $\vartheta g = \tau$, with the entry $\delta g = P$, and with the exit $\varrho g = N$ controlled by $\contr N = B$. Since there is usually at most one channel of a given type between two given configurations, we usually omit the label $g$, and write just $P \tto{\tau} N_B$ to denote this channel.

\subsection{Examples of networks}
\subsubsection{Cyber networks}\label{Cybnets}
Cyber networks are built following the \emph{``end-to-end``} architecture \cite{SaltzerJ:end-to-end}. In our formalism, a cyber network is characterized by the fact that 
\begin{itemize}
\item $\Chan = \{\mbox{cyb}\}$, i.e. there is just one channel type, which we call \emph{cyber} channel.  This is are the insecure channel over which cryptographic protocols are usually run.
\item $\Conf = \Nod$, i.e. the only configurations are the nodes.
\item There is a channel $M\to N$ for every pair $M,N\in \Nod$.
\item All communication is done by broadcast from the sender to all nodes in the network. The recipient does not observe the sender directly (although, of course, the sender can identify, or misidentify herself in the message). If a principal controls several nodes, it makes no difference which node he uses to send a message. Without loss of generality, we can thus assume that each principal controls exactly one node, i.e. that $ \Nod = \Ide$.
\end{itemize}
The actor-network structure of a cyber network is thus degenerate, and boils down to a single type $\Conf = \Nod = \Ide$, since the only configurations are the nodes, and the nodes are in one-to-one correspondence with the principals. That is why in crypto protocol analysis, we usually just specify how many different principals should play different roles. The fact that any two principals, \emph{viz} network nodes, are directly connected by a completely insecure cyber channel is assumed tacitly. The cyber network with three nodes/principals is presented on Fig~\ref{cybnet}.
\newcommand{\confA}{\mbox{\large $A$}}
\newcommand{\confB}{\mbox{\large $B$}}
\newcommand{\shake}{\mbox{\large $C$}}
\newcommand{\Mallory}{\mbox{\Large \textbf{\textit M}}}
\begin{figure}
\begin{minipage}[b]{0.4\linewidth}
\centering
\def\JPicScale{.75}
\ifx\JPicScale\undefined\def\JPicScale{1}\fi
\psset{unit=\JPicScale mm}
\psset{linewidth=0.3,dotsep=1,hatchwidth=0.3,hatchsep=1.5,shadowsize=1,dimen=middle}
\psset{dotsize=0.7 2.5,dotscale=1 1,fillcolor=black}
\psset{arrowsize=1 2,arrowlength=1,arrowinset=0.25,tbarsize=0.7 5,bracketlength=0.15,rbracketlength=0.15}
\begin{pspicture}(0,0)(54,40)
\rput(30,40){$\shake$}
\rput(6,2){$\confA$}
\rput(54,1){$\confB$}
\rput{0}(10,5){\psellipse[fillstyle=solid](0,0)(1.06,-1.06)}
\rput{0}(50,5){\psellipse[fillstyle=solid](0,0)(1.06,-1.06)}
\rput{0}(30,35){\psellipse[fillstyle=solid](0,0)(1.06,-1.06)}
\psline[linewidth=0.75,linecolor=gray,fillstyle=solid,arrowsize=1.5 2,arrowlength=2]{->}(27,33)(11,9)
\psline[linewidth=0.75,linecolor=gray,fillstyle=solid,arrowsize=1.5 2,arrowlength=2]{->}(32,32)(47,8)
\psline[linewidth=0.75,linecolor=gray,fillstyle=solid,arrowsize=1.5 2,arrowlength=2]{->}(49,9)(34,33)
\psline[linewidth=0.75,linecolor=gray,fillstyle=solid,arrowsize=1.5 2,arrowlength=2]{->}(13,8)(29,32)
\psline[linewidth=0.75,linecolor=gray,fillstyle=solid,arrowsize=1.5 2,arrowlength=2]{->}(46,6)(14,6)
\psline[linewidth=0.75,linecolor=gray,fillstyle=solid,arrowsize=1.5 2,arrowlength=2]{->}(15,4)(46,4)
\end{pspicture}
\caption{Cyberspace as completely connected}
\label{cybnet}
\end{minipage}
\hspace{1.3cm}
\begin{minipage}[b]{0.4\linewidth}
\centering
\def\JPicScale{.75}
\ifx\JPicScale\undefined\def\JPicScale{1}\fi
\psset{unit=\JPicScale mm}
\psset{linewidth=0.3,dotsep=1,hatchwidth=0.3,hatchsep=1.5,shadowsize=1,dimen=middle}
\psset{dotsize=0.7 2.5,dotscale=1 1,fillcolor=black}
\psset{arrowsize=1 2,arrowlength=1,arrowinset=0.25,tbarsize=0.7 5,bracketlength=0.15,rbracketlength=0.15}
\begin{pspicture}(0,0)(54,40.62)
\rput(30,40.62){$\shake$}
\rput(6,2){$\confA$}
\rput(54,1){$\confB$}
\rput{0}(10,5){\psellipse[fillstyle=solid](0,0)(1.06,-1.06)}
\rput{0}(50,5){\psellipse[fillstyle=solid](0,0)(1.06,-1.06)}
\rput{0}(30,36){\psellipse[fillstyle=solid](0,0)(1.06,-1.06)}
\psline[linewidth=0.75,linecolor=gray,fillstyle=solid,arrowsize=1.5 2,arrowlength=2]{->}(26,18)(12,8)
\psline[linewidth=0.75,linecolor=gray,fillstyle=solid,arrowsize=1.5 2,arrowlength=2]{->}(32,15)(47,6)
\psline[linewidth=0.75,linecolor=gray,fillstyle=solid,arrowsize=1.5 2,arrowlength=2]{->}(48,8)(33,17)
\psline[linewidth=0.75,linecolor=gray,fillstyle=solid,arrowsize=1.5 2,arrowlength=2]{->}(13,6)(28,16)
\psline[linewidth=0.75,linecolor=gray,fillstyle=solid,arrowsize=1.5 2,arrowlength=2]{->}(31.25,20)(31.25,32.5)
\psline[linewidth=0.75,linecolor=gray,fillstyle=solid,arrowsize=1.5 2,arrowlength=2]{->}(29,33)(28.75,20)
\rput{0}(30,18){\psellipse[fillstyle=solid](0,0)(1.06,-1.06)}
\rput(30,10){$\Mallory$}
\end{pspicture}
\caption{Cyberspace as completely adversarial}
\end{minipage}
\label{cybmal}
\end{figure}

The fact that the cyber channels are insecure can be captured in the model of a cyber network by assuming that all traffic is routed through the attacker, i.e. that $A$lice, $B$ob and $C$arol are all linked with each other through $M$allory. In a sense, the attacker Mallory is the embodyment of cyberspace. This architecture is presented on Fig.~2.

\subsubsection{An actor-network for two factor authentication}\label{two-fac-net}
\renewcommand{\confA}{Q}
\renewcommand{\shake}{R}
\newcommand{\humanA}{\Human_A}
\newcommand{\humanB}{\Human_B}
\newcommand{\deviceA}{C_A}
\newcommand{\cardA}{S_A}
\newcommand{\deviceB}{C_B}
\renewcommand{\shake}{R}
\newcommand{\vis}{\mbox{\small  vis}}
\newcommand{\bio}{\mbox{\small key}}
\newcommand{\both}{\mbox{\small shk}}
\newcommand{\cyb}{\mbox{\small cyb}}
\newcommand{\Alice}{\Large \mbox{$A$}}
\newcommand{\Bob}{\Large \mbox{$B$}}
To mitigate phishing attacks, most online banks have rolled out two factor authentication. This means that they do not just verify that the user knows a password, but also something else --- which is the second authentication factor. This second factor often requires some additional network resources, besides the internet link between the customer and the bank. This is the first, quite familiar step beyond simple cyber networks.

In the simplest case, the bank authenticates the browser used to access the service, by leaving a persistent cookie. The server often also records some data about user's computer and network location. The user only notices this when she tries to access the bank from another location, or using another browser: she is then asked to go through a round of  ``mother's maiden name" type of challenge questions. A more interesting type of second factor are the single-use Transaction Authentication Numbers (TANs) that the server may generate. Initially, they were be predistributed on paper. Nowadays they are often sent to user's mobile phone in an SMS message when login is initiated. The user is thus authenticated as the owner of her mobile phone. The other way around, the server is authenticated to be in possession of user's phone number, which eliminates the general phishing attacks.

Some banks authenticate that the user is in possession of her smart card. The underlying actor-network is on Fig.~\ref{pervnet}. The user $A$lice controls her computer $C_A$ and her smart card $S_A$. She is also given a portable smart card reader $R$. She inserts the card in the reader to form the configuration $Q$. The reader is available to Alice, but any other reader would do as well. Configured into $Q$, the smart card and the reader verify that Alice knows the PIN, and then generates the login credentials, which Alice copies from $R$'s screen to her computer $C_A$'s keyboard, which forwards it to bank Bob's computer $C_B$.  The details of the authentication procedure will be analyzed later. 

In summary, the network thus consists of
\begin{itemize}
\item principals $\Ide = \{A,B\}$, 
\item nodes $\Nod = \{\humanA, \deviceA, \cardA, \shake, \deviceB\}$;
\item configurations $\Conf = \Nod \cup \{\confA\}$, where $\confA= \{\cardA, \shake\}$,
\item and the following six channels
\begin{itemize}
\item cyber channels $\deviceA \leftrightarrows \deviceB$ between $A$lice's and $B$ank's computers,
\item visual channel $\deviceA \to \humanA$ from $A$lice's computer to her human $\humanA$,
\item keyboard $\humanA \to \deviceA$ from $A$lice's human to her computer,
\item visual channel $\shake \to \humanA$ from the smart card reader to $A$lice's human,
\item keyboard $\humanA \to \shake$ from $A$lice's human to the card reader.
\end{itemize}

\end{itemize}
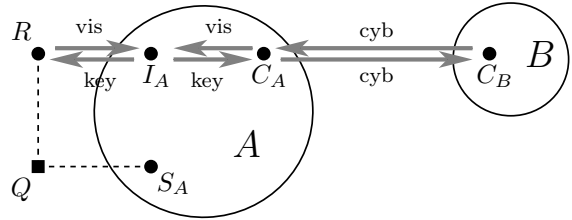
\begin{figure}
\begin{center}
\def\JPicScale{.75}
\ifx\JPicScale\undefined\def\JPicScale{1}\fi
\psset{unit=\JPicScale mm}
\psset{linewidth=0.3,dotsep=1,hatchwidth=0.3,hatchsep=1.5,shadowsize=1,dimen=middle}
\psset{dotsize=0.7 2.5,dotscale=1 1,fillcolor=black}
\psset{arrowsize=1 2,arrowlength=1,arrowinset=0.25,tbarsize=0.7 5,bracketlength=0.15,rbracketlength=0.15}
\begin{pspicture}(0,0)(99,39)
\rput(26,26){$\humanA$}
\rput(46,26){$\deviceA$}
\rput(2,34){$\shake$}
\rput(86,26){$\deviceB$}
\rput(2,6){$\confA$}
\psline[linestyle=dashed,dash=1 1](5,10)(25,10)
\psline[linestyle=dashed,dash=1 1](5,30)(5,10)
\rput{0}(85,30){\psellipse[fillstyle=solid](0,0)(1.06,-1.06)}
\rput{0}(25,10){\psellipse[fillstyle=solid](0,0)(1.06,-1.06)}
\rput{0}(45,30){\psellipse[fillstyle=solid](0,0)(1.06,-1.06)}
\rput{0}(25,30){\psellipse[fillstyle=solid](0,0)(1.06,-1.06)}
\rput{2.11}(34.31,19.76){\psellipse[](0,0)(19.37,-18.75)}
\psline[linewidth=0.75,linecolor=gray,fillstyle=solid,arrowsize=1.5 2,arrowlength=2]{->}(43,31)(29,31)
\psline[linewidth=0.75,linecolor=gray,fillstyle=solid,arrowsize=1.5 2,arrowlength=2]{->}(29,29)(43,29)
\rput(37,35){$\vis$}
\rput(35,25){$\bio$}
\rput(42,14){$\Alice$}
\rput(65,34){$\cyb$}
\rput(65,26){$\cyb$}
\rput{0}(5,30){\psellipse[fillstyle=solid](0,0)(1.06,-1.06)}
\psline[linewidth=0.75,linecolor=gray,fillstyle=solid,arrowsize=1.5 2,arrowlength=2]{->}(8,31)(23,31)
\rput(14,35){$\vis$}
\rput(29,7){$\cardA$}
\rput{0}(88.76,29){\psellipse[](0,0)(10.24,-10)}
\rput(94,30){$\Bob$}
\psline[linewidth=0.75,linecolor=gray,fillstyle=solid,arrowsize=1.5 2,arrowlength=2]{->}(82,31)(47,31)
\psline[linewidth=0.75,linecolor=gray,fillstyle=solid,arrowsize=1.5 2,arrowlength=2]{->}(48,29)(82,29)
\psline[linewidth=0.75,linecolor=gray,fillstyle=solid,arrowsize=1.5 2,arrowlength=2]{->}(22,29)(7,29)
\rput(16,25){$\bio$}
\pspolygon[fillstyle=solid](4,11)(6,11)(6,9)(4,9)
\end{pspicture}
\caption{A pervasive network: Online banking with a smart card reader}
\label{pervnet}
\end{center}
\end{figure}

\subsubsection{An actor-network for device handshake} \label{handshake-net}
\renewcommand{\confA}{Q_A}
\renewcommand{\confB}{Q_B}
\renewcommand{\deviceA}{\Device_A}
\renewcommand{\deviceB}{\Device_B}
\newcommand{\randomness}{S}
\renewcommand{\bio}{\mbox{\small bio}}
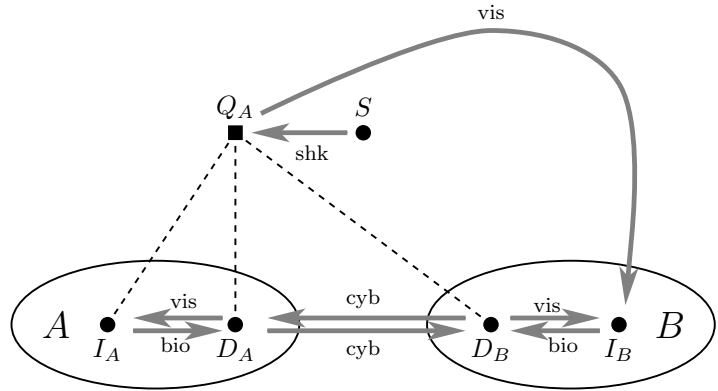
\begin{figure}
\begin{center}
\def\JPicScale{.85}
\ifx\JPicScale\undefined\def\JPicScale{1}\fi
\psset{unit=\JPicScale mm}
\psset{linewidth=0.3,dotsep=1,hatchwidth=0.3,hatchsep=1.5,shadowsize=1,dimen=middle}
\psset{dotsize=0.7 2.5,dotscale=1 1,fillcolor=black}
\psset{arrowsize=1 2,arrowlength=1,arrowinset=0.25,tbarsize=0.7 5,bracketlength=0.15,rbracketlength=0.15}
\begin{pspicture}(0,0)(105,59)
\rput(10,6){$\humanA$}
\rput(30,6){$\deviceA$}
\rput(70,6){$\deviceB$}
\rput(90,6){$\humanB$}
\rput(50,44){$\randomness$}
\rput(30,44){$\confA$}
\psline[linestyle=dashed,dash=1 1](30,40)(10,10)
\psline[linestyle=dashed,dash=1 1](30,40)(30,10)
\psline[linestyle=dashed,dash=1 1](30,40)(70,10)
\rput{0}(70,10){\psellipse[fillstyle=solid](0,0)(1.06,-1.06)}
\rput{0}(90,10){\psellipse[fillstyle=solid](0,0)(1.06,-1.06)}
\rput{0}(30,10){\psellipse[fillstyle=solid](0,0)(1.06,-1.06)}
\rput{0}(10,10){\psellipse[fillstyle=solid](0,0)(1.06,-1.06)}
\rput{0}(17.5,10){\psellipse[](0,0)(22.5,-10)}
\rput{0}(82.5,10){\psellipse[](0,0)(22.5,-10)}
\psline[linewidth=0.75,linecolor=gray,fillstyle=solid,arrowsize=1.5 2,arrowlength=2]{<-}(32.5,40)(47.5,40)
\pscustom[linewidth=0.75,linecolor=gray,arrowsize=1.5 2,arrowlength=2]{\psbezier{-}(34,43)(34,43)(60,56)(70,56)
\psbezier(80,56)(87.57,52.29)(90.29,47.15)
\psbezier(93,42)(92.64,31.04)(92,24)
\psline{->}(92,24)(91,13)
}
\psline[linewidth=0.75,linecolor=gray,fillstyle=solid,arrowsize=1.5 2,arrowlength=2]{->}(28,11)(14,11)
\psline[linewidth=0.75,linecolor=gray,fillstyle=solid,arrowsize=1.5 2,arrowlength=2]{->}(14,9)(28,9)
\psline[linewidth=0.75,linecolor=gray,fillstyle=solid,arrowsize=1.5 2,arrowlength=2]{->}(87,9)(73,9)
\psline[linewidth=0.75,linecolor=gray,fillstyle=solid,arrowsize=1.5 2,arrowlength=2]{->}(73,11)(87,11)
\rput(21.88,13.75){$\vis$}
\rput(78.75,13.12){$\vis$}
\rput(70,59){$\vis$}
\rput(81.25,6.88){$\bio$}
\rput(20.62,6.88){$\bio$}
\rput(98,10){$\Bob$}
\rput(2,10){$\Alice$}
\rput(42,37){$\both$}
\psline[linewidth=0.75,linecolor=gray,fillstyle=solid,arrowsize=1.5 2,arrowlength=2]{->}(66,11)(35,11)
\psline[linewidth=0.75,linecolor=gray,fillstyle=solid,arrowsize=1.5 2,arrowlength=2]{->}(35,9)(66,9)
\rput(50,14){$\cyb$}
\rput(50,6){$\cyb$}
\pspolygon[fillstyle=solid](29,41)(31,41)(31,39)(29,39)
\rput{0}(50,40){\psellipse[fillstyle=solid](0,0)(1.06,-1.06)}
\end{pspicture}
\caption{Actor-network for device handshake}
\label{quad-half}
\end{center}
\end{figure}
Suppose that Alice and Bob have some hand held devices and that they want to pair them, i.e. set up a secure cryptographic channel, without any previous encounters or infrastructure. There is a whole industry of methods to do this. We describe a method inspired by \cite{Mayrhofer:shake} and \cite{BuhanI:safe}. Alice's and Bob's devices $D_A$ and $D_B$ are equipped with accelerometers. If a device with an accelerometer is shaken, then the accelerometer can be used as a source of randomness. If two devices are shaken together, their accelerometers will generate roughly similar random strings. A shared random string can  be extracted by the techniques described in \cite{Mayrhofer:shake}, or using fuzzy extractors \cite{fuzzy-extractors}. We simply assume that a jointly samplable source is given, and denote id by the node $S$.

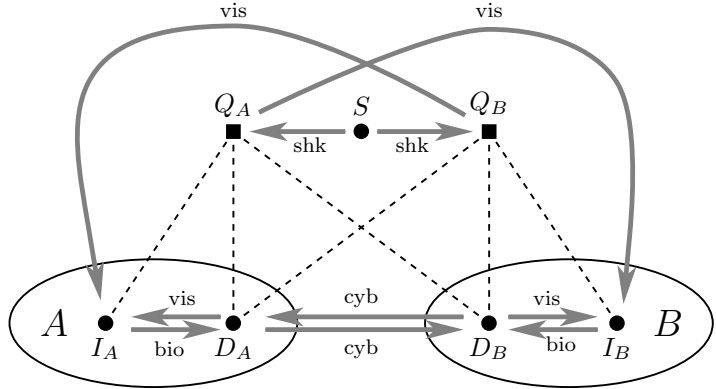
\begin{figure}
\begin{center}
\def\JPicScale{.85}
\ifx\JPicScale\undefined\def\JPicScale{1}\fi
\psset{unit=\JPicScale mm}
\psset{linewidth=0.3,dotsep=1,hatchwidth=0.3,hatchsep=1.5,shadowsize=1,dimen=middle}
\psset{dotsize=0.7 2.5,dotscale=1 1,fillcolor=black}
\psset{arrowsize=1 2,arrowlength=1,arrowinset=0.25,tbarsize=0.7 5,bracketlength=0.15,rbracketlength=0.15}
\begin{pspicture}(0,0)(105,59)
\rput(10,6){$\humanA$}
\rput(30,6){$\deviceA$}
\rput(70,6){$\deviceB$}
\rput(90,6){$\humanB$}
\rput(50,44){$\randomness$}
\rput(30,44){$\confA$}
\rput(70,44.38){$\confB$}
\psline[linestyle=dashed,dash=1 1](30,40)(10,10)
\psline[linestyle=dashed,dash=1 1](30,40)(30,10)
\psline[linestyle=dashed,dash=1 1](30,40)(70,10)
\psline[linestyle=dashed,dash=1 1](70,40)(70,10)
\psline[linestyle=dashed,dash=1 1](70,40)(30,10)
\psline[linestyle=dashed,dash=1 1](70,40)(90,10)
\rput{0}(70,10){\psellipse[fillstyle=solid](0,0)(1.06,-1.06)}
\rput{0}(90,10){\psellipse[fillstyle=solid](0,0)(1.06,-1.06)}
\rput{0}(30,10){\psellipse[fillstyle=solid](0,0)(1.06,-1.06)}
\rput{0}(10,10){\psellipse[fillstyle=solid](0,0)(1.06,-1.06)}
\rput{0}(17.5,10){\psellipse[](0,0)(22.5,-10)}
\rput{0}(82.5,10){\psellipse[](0,0)(22.5,-10)}
\psline[linewidth=0.75,linecolor=gray,fillstyle=solid,arrowsize=1.5 2,arrowlength=2]{<-}(32.5,40)(47.5,40)
\pscustom[linewidth=0.75,linecolor=gray,arrowsize=1.5 2,arrowlength=2]{\psbezier{-}(34,43)(34,43)(60,56)(70,56)
\psbezier(80,56)(87.57,52.29)(90.29,47.15)
\psbezier(93,42)(92.64,31.04)(92,24)
\psline{->}(92,24)(91,13)
}
\psline[linewidth=0.75,linecolor=gray,fillstyle=solid,arrowsize=1.5 2,arrowlength=2]{->}(28,11)(14,11)
\psline[linewidth=0.75,linecolor=gray,fillstyle=solid,arrowsize=1.5 2,arrowlength=2]{->}(14,9)(28,9)
\psline[linewidth=0.75,linecolor=gray,fillstyle=solid,arrowsize=1.5 2,arrowlength=2]{->}(87,9)(73,9)
\psline[linewidth=0.75,linecolor=gray,fillstyle=solid,arrowsize=1.5 2,arrowlength=2]{->}(73,11)(87,11)
\rput(22,14){$\vis$}
\rput(79,14){$\vis$}
\rput(70,59){$\vis$}
\rput(81.25,6.88){$\bio$}
\rput(20,6){$\bio$}
\rput(98,10){$\Bob$}
\rput(2,10){$\Alice$}
\rput(42,38){$\both$}
\psline[linewidth=0.75,linecolor=gray,fillstyle=solid,arrowsize=1.5 2,arrowlength=2]{->}(66,11)(35,11)
\psline[linewidth=0.75,linecolor=gray,fillstyle=solid,arrowsize=1.5 2,arrowlength=2]{->}(35,9)(66,9)
\rput(50,14){$\cyb$}
\rput(50,6){$\cyb$}
\pscustom[linewidth=0.75,linecolor=gray,arrowsize=1.5 2,arrowlength=2]{\psbezier{-}(66.24,42.54)(66.24,42.54)(43.24,56.54)(30.24,56.54)
\psbezier(17.24,56.54)(10.24,53.54)(7.24,47.54)
\psbezier(4.24,41.54)(6,33)(7.24,25.54)
\psline{->}(7.24,25.54)(9.24,13.54)
}
\rput(30,59){$\vis$}
\pspolygon[fillstyle=solid](29,41)(31,41)(31,39)(29,39)
\pspolygon[fillstyle=solid](69,41)(71,41)(71,39)(69,39)
\psline[linewidth=0.75,linecolor=gray,fillstyle=solid,arrowsize=1.5 2,arrowlength=2]{<-}(67.5,40)(52.5,40)
\rput(58,38){$\both$}
\rput{0}(50,40){\psellipse[fillstyle=solid](0,0)(1.06,-1.06)}
\end{pspicture}
\caption{Actor-network for two-round device handshake}
\label{quad}
\end{center}
\end{figure}

We now describe an actor-network supporting a \emph{device handshake procedure}, where the usual device pairing task is strengthened by the requirement that the secret shared by the devices $D_A$ and $D_B$ is also bound to the identities of their human owners $I_A$ and $I_B$. Fig.~\ref{quad-half} shows an actor-network supporting, in a sense, a half of this task: it will allow Alice to shake two devices together, to extract the shared secret; and \emph{moreover} Alice's device $D_A$ will biometrically verify that it is being shaken by Alice's human $I_A$, and not by someone else. If the device $D_A$ signals whether this verification succeeds in a way visible to Bob's human $I_B$, then Bob knows that the key extracted into his device $D_B$ is shared with $D_A$, \emph{and} bound to the identity of $I_A$. Alice's human $I_A$ can obtain similar assurances in an analogous round of the same procedure. The network for both rounds is depicted on Fig.~\ref{quad}.

Formally, in the actor-network for one round, depicted on Fig.~\ref{quad-half}, Alice controls the configuration $Q_A$, which consists of her device $D_A$, Bob's device $D_B$, and Alice's hand $I_A$, holding the two devices together. Alice's action of shaking the devices is represented as $Q_A$'s action of sampling a source of randomness $S$ along a devoted channel. The  accelerometers are abstracted away, and reduced to the node $S$. The fuzzy extractors are abstracted away and reduced to the fact that the randomness conveyed to $Q_A$ is distributed to both of the actors $D_A$ and $D_B$ participating in it. The details of the procedure are analyzed later. Here we just summarize that the actor-network for device handshake consists of the following data:
\begin{itemize}
\item two principals in $\Ide$, $A$lice and $B$ob, 
\item five nodes in $\Nod$:  
\begin{itemize}
\item $\deviceA$ and $\deviceB$ are Alice's and Bob's devices
\item $\humanA$ and $\humanB$ are Alice's and Bob's human identities,
\item $\randomness$ is a source of randomness;
\end{itemize}
\item one configuration in $\Conf$
\begin{itemize}
\item $\confA$ is the configuration where Alice holds her and Bob's devices in her hand: it consists of $\humanA$, $\deviceA$ and $\deviceB$;
\end{itemize}
\item eight channels in $\Chan$:
\begin{itemize}
\item two cyber channels $\deviceA \leftrightarrows \deviceB$ between $A$lice's and $B$ob's devices;
\item one biometric channel: 
\begin{itemize}
\item $\humanA\to \deviceA$ from Alice's hand to her device 
\end{itemize}

\item three visual channels: 
\begin{itemize}
\item  $\deviceA\to \humanA$ from Alice's device to her eyes,
\item  $\deviceB\to \humanB$ from Bob's device to his eyes,
\item $\confA \to \humanB$ from Alice's hand, holding both her and Bob's device to Bob's eyes;
\end{itemize}
\item one physical channel:
\begin{itemize}
\item $\randomness\to \confA$ from the source of randomness to Alice's configuration, which conveys the randomness to $D_A$ and $D_B$.
\end{itemize}
\end{itemize}
\end{itemize}
To assure that each principal contributes to the randomness, the above procedure can be repeated with  Bob shaking. The actor-network where both Alice and Bob see each other shaking both devices is depicted on Fig.~\ref{quad}

\subsection{Metaphysics of security}
\subsubsection{Knowing, Having and Being}
It is often repeated that security is based on:
\begin{itemize}
\item something you \emph{know}: digital keys, passwords, and other secrets,
\item something you \emph{have}: physical keys and locks, smart cards, tamper-resistant devices, or
\item something you \emph{are}: biometric features, such as fingerprints, eye irises, and other unmodifiable properties of your body; or the capabilities that you cannot convey to others, such as your handwriting.
\end{itemize}
An identity can thus be determined by its secrets, its tokens, and its features. In our model, this is captured by three levels of control that a principal may have over its nodes:
\begin{itemize}
\item \emph{secrets:} what you know can be copied and sent to others,
\item \emph{tokens:} what you have cannot be copied, but can be given away, whereas
\item \emph{features:} what you are cannot be copied, or given away.
\end{itemize}

\paragraph{Comments.} The common end-to-end security goals are usually realized by means of cryptographic software, and the principals prove their identities by their secrets. In cyber networks, \emph{a principal can be identified with the list of secrets that she knows}. If Alice and Bob share all their secrets, then there is no way to distinguish them by the challenges that can be issued on the standard completely insecure network channels\footnote{Here we assume that they share the secrets with each other dynamically: the secrets   it will immediately be shared with the other. This implies that they also observe the events on the same set of network nodes.}. For all purposes, they must be considered as the same principal. 

In pervasive networks, on the other hand, security is also supported by physical security tokens and hardware. Formally, this is where the network model becomes nontrivial: security tokens correspond to network nodes which may be controlled by one principal at one point in time, and  by another one at another point.  

Finally, security features correspond to network nodes which are controlled by a single principal, and cannot be relinquished. Such nodes correspond to biometric properties. Their counterpart are the nodes that represent  {\em biometric devices}. When all is well, a biometric channel thus has a biometric property at its entry, and a biometric device capable of observing this property at the exit.

\subsubsection{What is an identity?}
In a cyber network, anyone who knows all my secrets can impersonate me, and the standard models thus assume that an identity is a set of secrets.\footnote{In reality, even in an end-to-end network, two principals with the same set of secrets but, say, different computational powers, can be distinguished by timing their responses. Or they may be distinguished by their histories, since since may have derived their secrets from different initial data, as explained in \cite[Sec.~IV.D.1]{PavlovicD:SEFM10}. The standard models, however, abstract away all that.} In a pervasive network, even if someone knows all my secrets, we can still be distinguished as long as only one of us has my smart card, or my fingerprints, or if only one of us is standing at the door. In the actor-network model, besides the secrets, there are also the various tokens and features that a principal may control. An identity is thus a set of actors, which may include tokens and features. Formally, the set of principals $\Ide$ can be represented
\begin{itemize}
\item in a cyber network along the injection  
\bear 
\Knows & : &  \Ide \hookrightarrow \WP \TTt
\eear 
which assigns to each identity the set of terms that she knows \cite{PavlovicD:ESORICS06}; and
\item in an actor-network, along the injection 
\bear
\contrr \ :\  \Ide &  \hookrightarrow &  \WP \Chan\\
A & \mapsto & \{P\in \Chan\ |\ \contr P = A\}
\eear 
which assigns to each identity the set of configurations that she controls.
\end{itemize}
For simplicity, we assume here that the storage containing the terms from $\Knows(A)$ is subsumed among the nodes $\contrr A$. More about this in Sec.~\ref{MesAlg}. 

\paragraph{Why do we use the words ``principal" and ``identity" as synonyms?} For the benefit of the readers with a background in protocol analysis, let us emphasize that \emph{principals do not perform any actions} in the present model. The principals control their actors, and the actors perform the actions, and play roles. A principal determines if and when its actors perform an action, and can thus coordinate the order in which the actions of her actors will be executed. But without the actors, a principal cannot execute any actions. That is why the alternative term \emph{``identity"} may be preferred over \emph{``principal"}. On the other hand, for the special case of protocols, the concept still boils down to the familiar idea of principals, who play their roles in protocols, etc. So we retain that term as well.

\paragraph{Metaphysics of actors and principals.} Intuitively, the relation between a principal and her actors in an actor-network can be construed in terms of the \emph{mind-body duality}: the principal is the mind, and the actors are some parts of the body, that the mind can use to observe the world and to act in it. The data received or sampled by some actors through the suitable channels are directly available to the principal: e.g., the principal may control a camera, and observe the visual signal that the camera receives. On the other hand, some other actors may not convey their data to the principal that controls them: the camera may have no cable, or not enough light. The body has its limitations. Which type of information each actor conveys to its principal must be specified by the procedure specific axioms. Such specifications determine the semantics of each model and the intent of each actor-network procedure.

\subsubsection{What are the channel types?} 
Some of the channel types that we shall study  are:
\begin{itemize}
\item cyber channels: each node broadcasts to all nodes; there is no notion of distance; the recipient cannot observe the sender\footnote{We can assume that the sender always includes her identity into the message, within some standard format such as email. But such source claims can be easily spoofed in cyber space.} 
\item visual channel: the events at all nodes within some distance are observed; the observed nodes may or may not observe that they are observed;
\item binary channel: streams bits from one node to another, flipping them with some given probability. 
\end{itemize}
The binary channel is one of the basic concepts of information theory, capturing a simple notion of random noise. Intuitively, the cyber attacker can be viewed as \emph{``adaptive noise"}, disturbing the integrity of the messages.

\section{Actor-network processes}\label{Processes-sec}
\subsection{Computation and communication}
Computation in a network consists of \emph{events}, which are localized at nodes or configurations. An event that is controlled by a principal is an \emph{action}. 

Communication in a network consists of \emph{information flows} along the channels. Each flow corresponds to a pair of events:
\begin{itemize}
\item a \emph{write} event at the entry of the channel, and
\item a \emph{read} event at the exit of the channel.
\end{itemize}
There are two kinds of flows:
\begin{itemize}
\item \emph{messages}, which consist of a \emph{send action} at the entry of the channel, and a \emph{receive coaction} at the exit; and  
\item \emph{{\sample}s}, which consist of an \emph{{\probe} action} at the exit, and a \emph{{\appear} coaction} at the entry.
\end{itemize}
The information flows and the corresponding events are summarized in Table~\ref{events}. The black dots mark the actions. A consistent action-coaction pair is called an \emph{interaction}: i.e., an interaction consists of a send action and a receive coaction, or of a {\appear} coaction and a {\probe} action. We presently consider only these two types of interactions. Both the receive coactions and the emit coactions are construed as passive events: neither the principal who receives a message, nor the one who {\appear}s from a {\sample} controls when this happens. Of course, in reality a principal may, e.g. actively refuse to receive a message, or to {\appear} emit a {\sample}, etc. But our goal is to enable simple analyses, and we leave these details outside the scope of the general model.

\begin{table}
\begin{center}
\begin{oldtabular}{|c|c||c|c|}
\cline{3-4}
\multicolumn{2}{c||}{}& \multicolumn{2}{|c|}{events (actions$^\bullet$)} \\
\cline{3-4}
\multicolumn{2}{c||}{}& write &   read\\
 \hline \hline
\multirow{2}{1.2ex}{\begin{sideways}flows\end{sideways}} 
& message &   send$^\bullet$ &  receive \\
\cline{2-4}
& source & emit &   sample$^\bullet$ \\
 \hline
\end{oldtabular}
\end{center}
\caption{Flows and events}
\label{events}
\end{table}%

A computational process that is localized at a node proceeds as in the traditional models of computation. The node can be thought of as a state machine (e.g. a Turing machine), and the computational events change its state. An event at a configuration $P$ may changes the states of any of the actors $N\in P$. In the actual analyses, the state changes often need to be traced, but we did not encounter an example where an actual state machine would need to be specified. The most abstract models that capture the relevant features usually support the simplest analyses, hiding the implementation details.

Besides transferring information from one configuration to another, the flows also synchronize the events that take place at different localities, because:
\begin{itemize}
\item every receive coaction must be preceded by a corresponding send action, and 
\item every \probe\ action must be preceded by a corresponding \appear\ coaction.
\end{itemize}
If a {\sample} has not been {\appear}ted to anywhere, then there is nothing to \probe, and no {\probe}ion of that {\sample} can occur. If a message has not been sent, then the corresponding receive event cannot occur. So 
\begin{itemize}
\item when I receive a message, then I know that it must have been sent previously by someone; and 
\item when I {\probe} a {\sample}, then I know that someone must have {\appear}ed to this \sample . 
\end{itemize}
That is how I draw conclusions about non-local events from the observations of my own local actions. This is formalized in Sec.~\ref{orig-ax}.


\paragraph{Intuitions and ideas.} A configuration can be thought of as a mechanism, assembled from separate components that may be owned and controlled by different principals. Another view is that a configuration is like a team, composed of the players that came to act together towards some goal, but will separate and go their own ways when they are done. The usual (physical) handshake, confirming a social contact, can be viewed as a configuration. A couple dancing together is a configuration. A band playing music together is a configuration.

A channel is like a wire, connecting two configurations. The messages and the {\sample}s are two types of flows through a channel. If Alice wants to send a message, she needs a channel to send it on. Bob is on the other side of the channel, passively waiting to receive the message. If Bob wants to \probe\ a \sample, he needs a channel for that. Alice is on the other side of the channel, passively {\appear}ing. In both cases, Alice is at the entry of the channel, and Bob is at the exit of the channel.

Besides the channels that connect it to other configurations, a configuration may have internal methods for coordination among its nodes. E.g., a handshake is coordinated by two hands sensing each other. A couple of dancers develop signals that coordinate their dance. In some cases, such signals need to be made explicit as information flows. A configuration may need to send itself a message, or to sample itself as a source: e.g., to assure that genuine randomness is extracted. While the internal signaling and coordination capture controlled processes within a configuration, there are many processes that take place within a single configuration that are not entirely controlled. Dance groups and the bands of musicians have, besides the subtle forms of internal signaling, evolved complex external procedures to synchronize and coordinate. Such procedures can be thought of as primordial \emph{social software}. Analyzing them within a formal model might conceivably open interesting possibilities of electronic support.

\subsection{Formalizing data as terms}\label{MesAlg}
Each flow carries some data, which contain information. In abstract models, data are represented as terms of an algebra: the content of a message is an element of an algebra. We shall also represent the emission from a source as an element from the same algebra.  The algebraic operations correspond to the data processing operations. In the standard symbolic protocol model  \cite{Dolev-Yao}, the messages the terms of a free algebra of encryption and decryption operations. More general algebraic models allow additional operations, and additional equations \cite{CortierV:algebraic}. Recall that an algebraic theory is a pair $(O,E)$, where $O$ is a set of finitary operations (given as symbols with arities), and $E$ a set of well-formed equations (i.e. where each operation has a correct number of arguments) \cite{Gratzer}.

\be{defn}\label{mesalgdef}
An algebraic theory $\TTt = (O,E)$ is called a {\em data theory\/} if $O$ includes
a binary pairing $(-,-)$ operation, and the unary operations $\pi_1$ and $\pi_2$ such that $E$ contains the equations $
\pi_1(u,v) =  u$, $\pi_2(u,v) =v$, and
$\left(\left(x,y\right),z\right)  = \left(x,\left(y,z\right)\right)$.
A {\em data algebra} is a polynomial extension $\TTT[\Var]$ of a $\TTt$-algebra $\TTT$.
\ee{defn} 

\paragraph{Function notation.} When no confusion seems likely, we elide the function applications to concatenation, and write $f.x$ instead of $f(x)$. A function of two arguments $e(x,y)$ is thus identified identified with its curried form $e.x.y$, and $e.x$ abbreviates $e(x,-)$. By abuse of notation, the pair $(x,y)$ can thus be written as $x,y$, and $(x,(y,z)) = ((x,y),z)$ as $x,y,z$. 

When no confusion is likely, we even elide the dot from the concatenation and simply write $fx$ instead of $f.x$, or $f(x)$.

\paragraph{Tupling.} The equation $(x,(y,z)) = ((x,y),z)$ in the above definition implies that there is a unique $n$-tupling operation for every $n$. The first two equations imply that the components of any tuple can be recovered. 

\paragraph{Random values are represented by indeterminates.} A polynomial extension $\TTT[\Var]$ is the free $\TTt$-algebra generated by adjoining a set of {\em indeterminates\/} $\Var$ to a $\TTt$-algebra $\TTT$ \cite[\S8]{Gratzer}. The elements $x,y,z\ldots$ of $\Var$ are used to represent nonces and other randomly generated values. This is justified by the fact that indeterminates can be consistently renamed: nothing changes if we permute them. That is just the property required from the random values  generated in a run of a protocol.
Of course, this is not the only requirement imposed on nonces and random values: the other requirement is that they are known only locally, i.e. only by those principals who generate them, or who receive them in a readable message. This requirement is not formalized within the algebra of messages, but by the binding rules of process calculus \cite{PavlovicD:JCS05,PavlovicD:ESORICS06}. Here we capture it by the freshness axioms in Sec.~\ref{freshness-sec}.

\paragraph{Stores are nodes.} While the random values are thus algebraically presented as the indeterminates (i.e. as the variables in the polynomial extension), the stores (i.e. the variables used in computation) can be modeled as network nodes, each with a devoted read channel and a write channel. The property of such a node which is that it can store a value. The term stored in such a node determines its state. In this way, the usual notion of state, as a partial assignment of values to variables, is included within the network model. The state of a configuration is thus the product of the states of its actors, where some of the actors only task is to store some values.

\paragraph{Easy subterms.} We assume that every data algebra comes equipped with the \emph{easy subterm relation} $\sqsubseteq$. The idea is that that $s\sqsubseteq t$ implies that $s$ is a subterm of $t$ such that every principal who knows $t$ also knows $s$. In other words, the views $\Knows_A$  are lower closed under $\sqsubseteq$, as explained in \cite{PavlovicD:ESORICS06}. This is in contrast with hard subterms, which cannot be extracted: e.g., the plaintext $m$ and the key $k$ are hard subterms of the encryption $E.k.m$. In the Dolev-Yao  algebra, it is straightforward to define the easy subterm relation inductively. For general algebraic theories, the task of discerning the subterms gets complicated. A general treatment was attempted in \cite{PavlovicD:ESORICS06}.

\subsection{Formalizing events and processes}\label{proc-sec}

In this section we define processes, the events that processes engage in, and the ordering of events within a process.

\subsubsection{Events}
An event or action is generally written in the form $a[t]$  where 
\begin{itemize}
\item $a$ is the event identifier,
\item $t$ is the term on which the event may depend.
\end{itemize}
When an event does not depend on data, the term $t$ is taken to be a fixed constant $t = \checkmark$, and we often abbreviate $a[\checkmark]$ to $a$. 

The most important events for our analyses are the action-coaction couples send-receive, and \probe-\appear, for which we introduce special notations:
\begin{itemize}
\item send $\send t$, receive $\recv t$, 
\item \appear\ $\apr t$, \probe\ $\prb t$.
\end{itemize}
Generically, we write
\begin{itemize}
\item $\writ t$ for a write action, which can be either $\send t$ or $\apr t$, and
\item $\reaa t$ for a read action, which can be either $\recv t$ or $\prb t$.
\end{itemize}
Another often used action is
\begin{itemize}
\item generate a random value $\nu[x]$,
\end{itemize}
It could also be implemented as sampling a source of randomness represented as a devoted node.

In addition, the nodes are capable of performing various local operations. Most are able to execute the standard pseudo-code commands, like comparisons $(t=s)$ or assignments $(t:= s)$. But the differences in their computational resources will be essential in some of the security analyses of the procedures below. Further examples of application specific events and actions will be introduced in the below.

For actions, such as $\send t$ and $\prb t$, the configuration $P$ must be controlled, i.e. the partial function $\contr:\Nod \rightharpoonup \Ide$ must have a definite value $\contr P$.

\paragraph{Representing events as terms.} How do we represent principals' observations of events and of other principals' actions? The location of an event   may be viewed as a source for sampling; the location of an action must be controlled by a principal, who may be viewed as the sender of the message about  the action that took place. But since only data can be sent as messages, or sampled from sources, each observable action $a[t]_P$ must be represented as a term $\Big[a[t]_P\Big]$. In general, this is done by adding to the presentation of the algebra $\TTt$ a mapping 
\bear
\trm{-} & : &  \EEe \to \TTt
\eear
which generates a representation of each event from $\EEe$. A similar map assigning to each action a logical formula leads to dynamic logic \cite{Kozen:dynlog-book}. We'll see a typical example of a message about an action in Sec.~\ref{handshake-proc}, where we return to the  device handshake procedure. Bob can only conclude that the secret shared by his and Alice's device is authentic if he sees Alice shaking the devices. 

\paragraph{Self-sampling.} A less typical, but not less interesting example arises if we assume that a configuration has a channel to sample its own events. Such a configuration can then sample its own sampling, i.e. execute the action $\prb{\trm{\prb{}}}$. The principal who controls such a configuration can then observe some of her own observations. It is interesting to explore the authenticity of such observations. We shall touch this again in Sec.~\ref{authchans-sec}.

\subsubsection{Processes}
\be{defn}\label{proc-def}
A \emph{process} $\FFF$ is a partially ordered multiset of localized events, i.e. a mapping 
\bear
\FFF\  =\  <\FFF_\EEe, \FFF_\Conf> & : &  \FFf\to \EEe\times \Conf
\eear
where
\begin{itemize}
\item $(\FFf,\rar)$ is a well-founded partial order, representing the structure time,
\item $\EEe$ is a family of events, and
\item $(\Conf, \subseteq)$ the partial order of configurations,
\end{itemize}
and they satisfy the requirements that
\begin{anumerate}
\item if $\FFF_\EEe \phi$ is an action, then $\contr ( \FFF_\Conf \phi)$ is well defined, and
\item if $\phi \rar \psi$ in $\FFf$ then $\FFF_\Conf \phi \subseteq \FFF_\Conf \psi$ or $\FFF_\Conf \phi \supseteq \FFF_\Conf\psi$ in $\Conf$.
\end{anumerate}
\ee{defn}

\paragraph{Notation: The points in time are denoted by events.} By abuse of notation, we usually write $a[t]_P$ for $\phi\in \FFF$ where $\FFF_\EEe\phi = a[t]$ and $\FFF_\Conf= P$. Of course, if the there are several points in time $\phi_1, \phi_2,\ldots \in \FFF$ where the same $P$ executes the same $a[t]$, then this notation is ambiguous, since it is not clear to which $\phi_i$ does $a[t]_P$ refer. But such situations are rare.  On the other hand, with this notation the above conditions become:
\begin{anumerate}
\item if an action takes place at a configuration $P$, then $P$ is controlled, i.e. $\contr P$ must be well defined, and
\item if $a[t]_P \rar b[s]_Q$ then $P \subseteq Q$ or $P \supseteq Q$.
\end{anumerate} 

\paragraph{Remarks.} The subset ordering of $(\Conf, \subseteq)$ arises from Def.~\ref{AN-def}, which says  that configurations are finite sets. Partially ordered multisets, or \emph{pomsets} were introduced and extensively studied by Vaughan Pratt and his students \cite{PrattV:Pomsets}.  Condition (a) specifies what we already said informally: that the configuration where an action takes place must be controlled. Condition (b), on the other hand, means that \emph{there is no subliminal synchronization}: the ordering of events can  only be imposed within a configuration that enables all of them. If Alice performs one action controlling a configuration $P_A$, and then another action controlling a configuration $Q_A$, then she must control a configuration $R_A\supseteq P_A\cup Q_A$, that will allow her to control the order of the actions with $P_A$ and with $Q_A$. The intended use of configurations, including any constraints that would require that some parts of $P_A$ and $Q_A$ should not be used together, must be imposed through axioms, in the PDL language introduced in Sec.~\ref{Logic-sec}.

%
\begin{defn}\label{orig-def}
We say that the term $t$ \emph{originates} at the point $\phi\in \FFF$ if $\phi$ is the earliest write of a term containing $t$. Formally, $\phi$ thus satisfies
\begin{itemize}
\item $\FFF_\EEe \phi = \writ s$ where $t\sqsubseteq s$, and
\item $\FFF_\EEe \xi = \writ s \wedge t\sqsubseteq s \then \phi \rar \xi$ holds for all events $\xi$.
\end{itemize}
\ee{defn}

\paragraph{Notation: Origination.} We extend the notational conventions described above by denoting by $\orig{\wwrit{t}_P}$ the event $\phi$ where the term $t$ originates. The configuration $P$ is the \emph{originator} of $t$.

\subsection{Formalizing flows, runs and procedures}

We now extend our discussion to the definition of communication between processes, and extend our ordering to events occurring within a procedure as well as individual processes.

We begin by defining a more general version of channel between two configurations, called a flow channel.  A flow channel exists between any two configurations if
a channel exists between any two nodes on the configuration trees.  It is called a flow channel because the information passed along the channel flows upwards to the
configuration as a whole.  It is defined formally below.

\be{defn}\label{flow-def}
For configurations $P,Q\in \Conf$, a \emph{flow channel} $P\flow \tau Q$ can be either
\begin{itemize}
\item a channel $P \tto \tau Q$, or
\item a flow channel $P \flow \tau Q'$, where $Q'\in Q$, or
\item a flow channel $P' \flow \tau Q$, where $P'\in P$, or
\item a flow channel $P' \flow \tau Q'$, where $P'\in P$ and $Q'\in Q$.
\end{itemize}


A \emph{flow} $a[t]_P \flow \tau b[s]_Q$ is given by
\begin{itemize}
\item a flow channel $P\flow \tau Q$, and
\item an interaction pair $a[t], b[s]$, i.e. a pair where
\begin{itemize}
\item either $a[t] = \send{t}$ and $b[s] = \recv{s}$, 
\item or $a[t] = \apr{t}$, and if $b[s] = \prb{s}$.
\end{itemize}
\end{itemize}

A flow $a[t]_P \flow \tau b[s]_Q$ is \emph{complete} if $s=t$.
\ee{defn}

\be{defn} \label{run-def}
Let $\FFF$ be a process. A \emph{run}, or \emph{execution} $\EEE^\FFF$ of $\FFF$ is an assignment for each coaction $b[s]_Q$ of a unique  flow $a[t]_P\flow{\tau} b[s]_Q$, which is required to be \emph{sound}, in the sense that $b[s]_Q\not\shortrightarrow a[t]_P$ in $\FFF$.

A run is \emph{complete} if all of the flows that it assigns are complete: the terms that are received are just those that were sent, and the inspections find just those terms that were submitted.
\ee{defn}

\paragraph{A run is a pomset extending its process.} Setting $a[t]_P \rar b[s]_Q$ whenever there is a flow $a[t]_P \flow{\tau} b[s]_Q$ of some type $\tau$ makes a run $\EEE^\FFF$ into an extension of the ordering of the process $\EEE$, as a partially ordered multiset. The pomset $\EEE^\FFF$ does not have to satisfy condition (b)  of Def.~\ref{proc-def} any more. Indeed, the whole point of running a process is to extend in $\EEE^\FFF$ the internal synchronizations, given by the ordering of $\FFF$, with the additional external synchronizations.

\paragraph{Overloading arrows.} The view of the runs as order extensions of the processes justifies the overloading of the arrow notation, which is used both 
\begin{itemize}
\item as $a[t]_P\rar b[s]_Q$, saying that $a[t]_P$ precedes $b[s]_Q$ in the partial ordering $(\FFF,\rar)$, and
\item $a[t]_P\flow{\tau} b[s]_Q$, denoting a flow of type $\tau$ from $a[t]_P$ to $b[s]_Q$ in a run $\EEE^\FFF$.
\end{itemize}
This overloading is consistent, because a flow from $a[t]_P$ to $b[s]_Q$ implies that $a[t]_P$ precedes $b[s]_Q$; and it will be useful when we pass from the runs, e.g. in Figures~6--8, to formal reasoning about them in Figures~9--12. The arrows in the latter family of diagrams arise from the arrows in the former family. But the former represent reality, whereas the latter represent assertions about it.

\be{defn}
A \emph{network procedure} $\LLL$ is a pair $\LLL=<\FFF_\LLL, E_\LLL>$ where 
\begin{itemize}
\item $\FFF_\LLL$ is a process, and
\item $E_\LLL = \{\EEE_1^{\FFF_\LLL}, \EEE_2^{\FFF_\LLL}, \EEE_3^{\FFF_\LLL}\ldots\}$ is a set of runs of $\FFF_\LLL$. 
\end{itemize}
The elements of $E_\LLL$ are called \emph{secure} runs. All other runs are \emph{insecure}. A procedure is said to be secure if every insecure run can be detected by a given logical derivation from the observations of a specified set of participants.
\ee{defn}

\paragraph{Procedures generalize protocols.} A \emph{protocol} is a special case of a network procedure, where the underlying network is a cyber network. Since cyber channels offer no security guarantees, the security goals of protocols are generally realized by cryptographic functions computed at the nodes. That is why cyber security is largely concerned with \emph{cryptographic} protocols. It is thus based on the \emph{end-to-end} paradigm, where the security tasks are pushed to the local computations at the smart ``ends", i.e. nodes, leaving the network simple and efficient.  The above definition of a secure procedure generalizes the definition of  a secure protocol used in Protocol Derivation Logic \cite{PavlovicD:ESORICS04,PavlovicD:ESORICS06,PavlovicD:MFPS10}, as well as in Protocol Composition Logic \cite{PavlovicD:JCS04,PavlovicD:JCS05,PCL}. Network procedures and their security proofs thus extend cryptographic protocols, and their security proofs. What is the difference? First of all, any node in a cyber network is as good as any other node, so it does not matter which ones you control, or how many. Without loss of generality, configurations can thus be reduced to single nodes, and the channel flows to the messages on cyber channels. A protocol run thus boils down to a sequence of messages among the principals, each usually controlling a single node, with some local computations in-between the messages.

Nontrivial configurations arise in pervasive networks. E.g., a smart card can only compute when inserted into a correct configuration with a card reader. Moreover, the information can flow through a pervasive network in many different ways: by messages sent along a variety of different channels, short range, cellular, social, etc.; or by observations along visual channels, etc. The distinction between computation and communication in pervasive networks becomes blurred, as the two become intertwined in subtle and complicated ways.

\paragraph{Graphic presentations of procedures.} To specify a procedure $\LLL$, we draw a picture of the pomset $\FFF= \FFF_\LLL$, and then each of its extensions $\EEE = \EEE^{\FFF_\LLL}_i$. Because of condition (b) of Def.\ref{proc-def}, the events comparable within the ordering of a process $\FFF$ must happen within a maximal configuration. Therefore, if the diagram of the partially ordered multiset $\FFF$ is drawn together with the underlying network, then each component of the comparable events can all be depicted under the corresponding configuration. We can thus draw the network above the process, and place the events occurring at each configuration along the imaginary vertical lines flowing, say, downwards from it, like in Fig.~\ref{cap-run}. The additional ordering, imposed when in a run $\EEE$ the messages get sent and the facts get observed, usually run across, from configuration to configuration. This ordering can thus be drawn along the imaginary horizontal lines between the events, or parallel with the channels of the network. Such message flows can also be seen in Fig.~\ref{cap-run}. The dashed lines represent the data sharing within a configuration.

This discipline of drawing 
\begin{itemize}
\item the internal ordering of events along the verticals and \item the external ordering, imposed by the flows, along the horizontal lines is
\end{itemize}
of course, familiar from strand spaces, where the verticals are the strands, and the horizontals the bundles \cite{strands}. Our diagrams indeed boil down to strand diagrams whenever the network configurations are single nodes connected by cyber channels, and when the only flows are messages.  This graphic convention for depicting the internal and external ordering of events goes back to the early days of distributed computing research, see e.g. \cite{Lamport:Time,McCullough:hookup}.

\subsection{Examples of procedures}
%
%
%

\subsubsection{Challenge response authentication protocols}\label{CR-prot}

\renewcommand{\deviceB}{B}
\renewcommand{\humanA}{A}
\renewcommand{\cyb}{\mbox{\small cyb}}
\newcommand{\screen}{\circ}
\newcommand{\respcr}{\scriptstyle r:=r^{AB} x}
\newcommand{\qm}{\mbox{\footnotesize $c^{AB}x$}}
\newcommand{\ru}{\mbox{\footnotesize $r$}}
\newcommand{\newn}{\scriptstyle \nu[x]}
\begin{figure}
\begin{center}
\def\JPicScale{.75}
\input{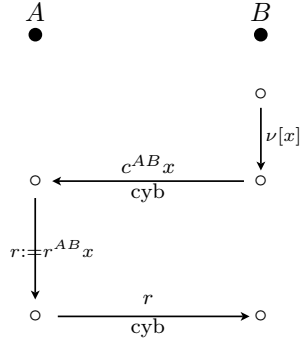}
\caption{Challenge-Response (CR) protocol template}
\label{CR-run}
\end{center}
\end{figure}
We begin a familiar special case of a procedure: a protocol. A large family of challenge-response authentication protocols is subsumed under the template depicted on Fig.~\ref{CR-run}. Bob wants to make sure that Alice is online. It is assumed that Alice and Bob share some sort of a secret $k^{AB}$, which allows them to define functions $c^{AB}$ and $r^{AB}$ such that 
\begin{itemize}
\item $r^{AB}x$ can be computed from $c^{AB} x$ using $s^{AB}$, but
\item $r^{AB}x$ can\emph{not} be computed from $c^{AB} x$ alone, without $s^{AB}$.
\end{itemize}
So Bob generates a fresh value $x$, sends the challenge $c^{AB} x$, and if he receives the response $r^{AB} x$ back, he knows that Alice must have been online, because she must have originated the response. The idea behind this template has been discussed, e.g., in \cite{PavlovicD:ESORICS04,PavlovicD:CSFW05,PavlovicD:ESORICS06,PavlovicD:MFPS10}. The template instantiates the concrete protocol components by refining the abstract functions $c^{AB}$ and $r^{AB}$ to concrete implementations, which satisfy the above requirements: e.g., $c^{AB}$ may be the encryption by Alice's public key, and $r^{AB}$ may be the encryption by Bob's public key, perhaps with Alice's identity.

Recall from in Sec.~\ref{Cybnets} that cyber networks are degenerate, in the sense that the actors boil down to the principals. Alice's and Bob's unique actors are thus simply denoted by $A$ and $B$.

\subsubsection{Two-factor authentication procedure}\label{two-fac-proc}
Next we describe the first nontrivial procedure, over the actor-network described in Sec.~\ref{two-fac-net}. It can be viewed as an extension of the simple challenge-response authentication. There, Bob authenticates Alice using her knowledge of  a secret $s^{AB}$, which they both know. Here Bob authenticates that that knows a secret $p^{A}$ that Bob does not know, and that she has a security token $S_A$, in this case a smart card. The secret and the smart card are the ``two factors``.  This is the idea of the procedure standardized under the name \emph{Chip Authentication Programme (CAP)}, analyzed in \cite{AndersonR:CAP}.  The desired run of the challenge-response option of this procedure is depicted on Fig.~\ref{cap-run}. 

We assume that, prior to the displayed run, Alice the customer identified herself to Bob the bank, and requested to be authenticated. Bob's computer $C_B$ then extracts a secret $s^{AB}$ that he shares with Alice. This time, though, the shared secret is too long for Alice's human $I_A$ to memorize, so it is is stored in the smart card $S_A$. Just like in CR protocol above, Bob issues a challenge, such that the response can only be formed using the secret. So Bob in fact authenticates the smart card $S_A$. He entrusts the smart card $S_A$ with authenticating Alice's human $I_A$. This is done using the secret $p^A$ shared by $I_A$ and $S_A$. The secret is stored in both nodes. To form the response to Bob's challenge, Alice forms the configuration $Q$ by inserting her card $S_A$ into the reader $R$. The configuration $Q$ requests that $I_A$ enters the secret PIN (Personal Identification Number) $p^A$ before it forms the response for Bob. There is no challenge from $Q$ to $I_A$, and thus no freshness guarantees in this authentication: anyone who sees $I_A$'s response can replay it at any time. Indeed, the human $I_A$ cannot be expected to perform computations to authenticate herself: most of us have trouble even submitting just the static PIN. The solution is thus to have the card-reader configuration $Q$ computes the response, which Alice relays it to Bob. The old PIN authentication is left to just convince $Q$ that Alice's human $I_A$ is there: $Q$ tests $p^A$, sent through the keybord channel from $\Human_A$ to the reader $R$, coincides with $\overline p^A$ stored in the card $S_A$, and then generates a keyed hash $Hs^{AB}x$ using the shared secret $s^{AB}$ and the challenge $x$. This hash is displayed for Alice on the card reader $\shake$ as the response $r$, which Alice then sends to her computer $C_A$ by the keyboard channel, and further to $C_B$ by the cyber channel.

This two-factor procedure is thus more secure than the simple password authentication of $\Human_A$ to $\Device_B$ because
\begin{itemize}
\item there is a fresh challenge, and the attacker cannot impersonate Alice just by recording one session (like phishermen do);
\item even in the option without the challenge, the secret $s^{AB}$, shared between two computing devices, is generally stronger than a human memorable secret $p^A$, and finally
\item the PIN authentication to the smart card is not cryptographically strong, but it is done on a physical channel, which is harder to attack.
\end{itemize}

If a thief comes in the possession of the smart card $S_A$, he cannot use it without $p^A$, stored in $\Human_A$. This leads to the muggings, i.e. attacks where $S_A$ is stolen, and then $\Human_A$ is coerced to disclose the PIN $p^A$. The authors of \cite{AndersonR:CAP} point out that introducing the portable, generic readers $R$ simplifies for the attacker the verification that the number given to him by $\Human_A$ is indeed the correct $p^A$. 

\renewcommand{\confA}{Q}
\renewcommand{\deviceA}{C_A}
\renewcommand{\deviceB}{C_B}
\renewcommand{\humanA}{I_A}
\renewcommand{\vis}{\mbox{\small vis}}
\renewcommand{\bio}{\mbox{\small kyb}}
\renewcommand{\cyb}{\mbox{\small cyb}}
\renewcommand{\screen}{\circ}
\renewcommand{\qm}{\mbox{\footnotesize $x$}}
\newcommand{\pwd}{\mbox{\footnotesize $p^A, x$}}
\newcommand{\ident}{\mbox{\footnotesize $s^{AB}$}}
\newcommand{\hash}{\mbox{\footnotesize  $r := Hs^{AB}x$}}
\newcommand{\cardpin}{\mbox{\small $\overline p^A$}}
\newcommand{\checkk}{\mbox{\footnotesize $p^A =\overline p^A$}}
\newcommand{\resp}{\mbox{\footnotesize $r$}}
\renewcommand{\ru}{\mbox{\footnotesize $r$}}
\renewcommand{\newn}{\scriptstyle \nu[x]}
\newcommand{\OK}{\scriptstyle r = Hs^{AB}x}
\begin{figure}
\begin{center}
\def\JPicScale{.75}
\input{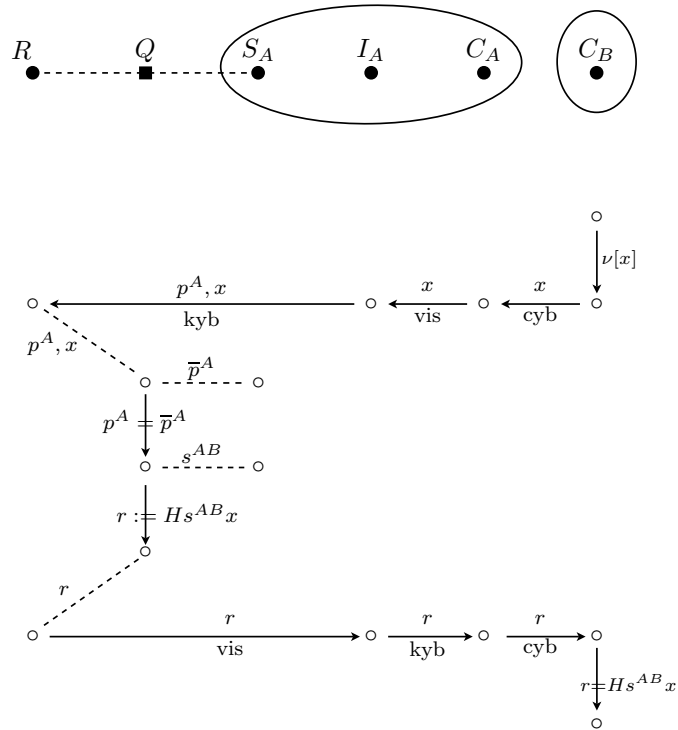}
\caption{Chip Authentication Program (CAP)  procedure}
\label{cap-run}
\end{center}
\end{figure}

\subsubsection{Device handshake procedure} \label{handshake-proc}
\renewcommand{\confA}{Q_A}
\renewcommand{\bio}{\mbox{\small bio}}
Going back to Fig.~\ref{quad-half}, we describe the desired run on the Device Handshake. The run begins by Alice bringing together the nodes for the configuration $\confA$. This means that she takes her device $D_A$ and Bob's device $D_B$ into her hand $I_A$, to shake them together.  We also assume that the configuration contains the node $S$, which is the source of randomness. This node does not correspond to a physical object, but embodies the source that is being sampled by shaking. In reality, the two devices (i.e. their accelerometers) record the joint action of shaking together, and sample the shared secret out of it. The method to achieve this is described in \cite{Mayrhofer:shake}, and we take it as given. The desired run on the network from Fig.~\ref{quad-half} is depicted on Fig.~\ref{DH-run}. It consists of five flows, triggered by $\confA$'s shaking of the devices:

\newcommand{\newx}{\scriptstyle \nu[x]}
\newcommand{\sendx}{\scriptstyle x}
\newcommand{\recvx}{\scriptstyle \checkmark}
\newcommand{\rrecvx}{\scriptstyle \trm{\prb{}}}
\newcommand{\one}{\scriptstyle {\rm ok}}
\newcommand{\finger}{\scriptstyle f^A}
\renewcommand{\humanA}{A}
\newcommand{\shk}{\mbox{\footnotesize shk}}
\renewcommand{\respcr}{\scriptstyle r:=r^{AB} x}
\renewcommand{\qm}{\mbox{\footnotesize $c^{AB}x$}}
\renewcommand{\ru}{\mbox{\footnotesize $r$}}
\renewcommand{\deviceB}{D_A}
\renewcommand{\deviceA}{I_A}
\renewcommand{\confA}{Q_A}
\renewcommand{\cardA}{D_B}
\renewcommand{\humanA}{I_B}
\renewcommand{\shake}{S}
\begin{figure}
\begin{center}
\def\JPicScale{.75}
\input{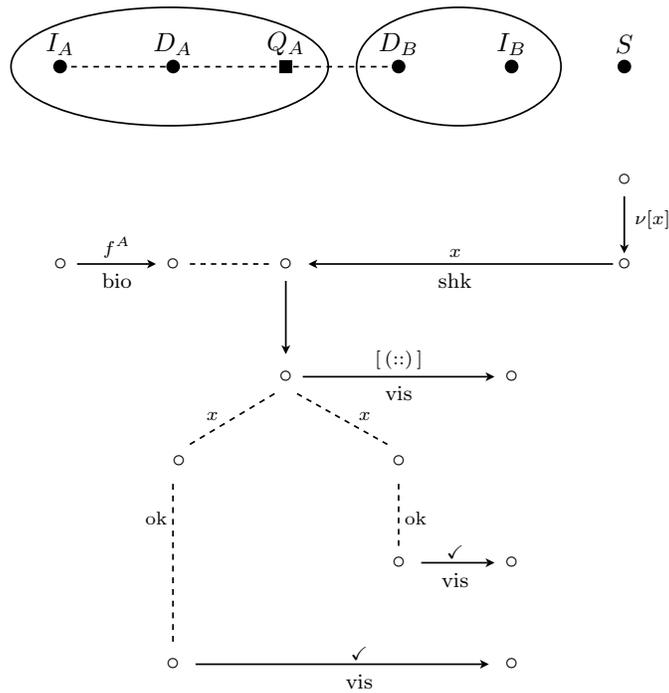}
\caption{Device handshake procedure}
\label{DH-run}
\end{center}
\end{figure}

\begin{itemize}
\item $\prb{x}_{\confA} \oot{\both} \apr{x}_\randomness$, caused by $\confA$'s sampling of $x$ from $S$;
\item $\prb{f^A}_{D_A} \oot{\bio} \apr{f^A}_{I_A}$, simmultaneously caused by $\deviceA$'s sampling of $I_A$'s fingerprint $f^A$; 
\item $\send{\trm{\prb{}_{Q_A}}}_{\confA} \tto {\vis}\recv{\trm{\prb{}_{Q_A}}}_{\humanB}$, where $\confA$ shows $\humanB$ how she shakes the devices to sample $x$;
\item $\send{\checkmark}_{D_A} \tto{\vis} \recv{\checkmark}_{I_B}$, confirming to $I_B$ that $D_A$ received $x$ and correct $f^A$; and
\item $\send{\checkmark}_{D_B} \tto{\vis} \recv{\checkmark}_{I_B}$, confirming to $I_B$ that $D_B$ received $x$.
\end{itemize}

\paragraph{Remark.} Note that the third flow contains an example of a message about an action, along the lines explained in Sec.~\ref{proc-sec}, in the paragraph about representing events as terms. Here Bob's human $I_A$ receives from Alice's configuration $Q_A$ the visual message $\trm{\prb{}}$ that she has sampled the randomness.

\paragraph{Two-round device handshake.} If both Alice and Bob need to be assured that a secret is shared, then the last two messages, where $D_A$ and $D_B$ confirm that they succeeded to sample $x$, and one of them confirms that the fingreprint is correct, should be sent to both $I_A$ and $I_B$. Running two rounds of the device handshake, with both Alice and Bob shaking their devices in the actor-network from Fig.~\ref{quad}, would require such procedure in each of the two rounds.

\section{Procedure Derivation Logic}\label{Logic-sec}
\subsection{Procedures are distributed predicates}
Formal methods for software engineering have been built upon Hoare's slogan that \emph{``Programs are predicates"} \cite{Hoare:ProgPred}. The view that computations can be adequately approximated by their logical descriptions, and proven correct, was the guiding principle of theory and practice of software specifications from the outset in the 1960s \cite{Floyd}.
 The underlying assumption is that software is run within a single computer, with a global observer asserting the predicates. In contrast, the essence of a network is that there is no global observer. The events at each node may be directly observed only by the principal who controls that node. In addition, they may be indirectly observed along an authentic channel that ends at that node.

The upshot is that each participant of network computation may observe different events, and assert different predicates. Each of them sees just bits and pieces of the network process. In most cases of interest, their common goal is to coordinate their observations, and arrive at a coherent joint view of the events in which they jointly participate, so that they can all assert the same predicate, which is the required security property. Towards this goal, they use the channels between them to exchange messages, and to observe each other. 

In summary, our goal is thus to extend formal methods from programs as predicates to network procedures as families of predicates, asserted at different network nodes. The task of reconciling these local assertions is usually formalized as \emph{authentication}. In this section, we gradually introduce the language and the axioms that allow participants of network computations to annotate their local observations by global predicates, building up from the basic communication axioms, towards authentication, and beyond.

\subsection{The language of PDL}
A statement of PDL is in the form $A:\Phi$, where $A\in \Ide$ is a principal, and $\Phi$ is a predicate asserted by $A$. The predicate $\Phi$ is formed by applying logical connectives to the atomic predicates, which can be
\begin{itemize}
\item $a[t]_P$ --- meaning ``the event $a[t]_P$ happened"; or
\item $a[t]_P\rar b[s]_Q$ --- meaning ``the event $a[t]_P$ happened before $b[s]_Q$".
\end{itemize}

\paragraph{Notation: Statements assert events, and events describe points of time.} Recall that the expressions like $a[t]_P$ refer to points in time descriptively, i.e. by specifying what happened and where. As explained in Sec.~\ref{proc-sec}, this is a notational abuse, but an important one. The descriptions $a[t]_P$ are sometimes refined to $\orig{a[[t]]_{P}}$, which say that the event $a[s]$ took place at the configuration $P$, for some $s\sqsupseteq t$, \emph{and} that the term $t$ originates there.

\paragraph{The essence of PDL.} Here, the event descriptions like $a[t]_P$ and $\orig{a[[t]]_{P}}$ moreover denote the assertions that the described event took place. The descriptions of processes and of their runs are used as the predicates to annotate them, and to reason about them. This notational abuse is justified by the \emph{isomorphism of process executions and their logical annotations}.  This is the basic design decision on which PDL is based: \emph{The descriptions of processes and of their runs as partial orders are used as the logical assertions to annotate these processes and their runs.} This is the formal implementation of the guiding principle of PDL, explained in Sec.~\ref{PCL-PDL}.

This isomorphism does not only simplify notation, but substantially simplifies the logic that we work with. A statement in any protocol or procedure logic is an assertion of a participant --- \emph{made at a certain point of a run}. In PCL, this dependency on the run was expressed using dynamic modalities. In PDL, though, the process expression within a modality isomorphic to a logical formula. So instead of 
\begin{itemize}
\item $A: [\psi]\Phi$, saying that $A$ knows that $\Phi$ is valid after the execution point $\psi$ is reached, we can write 
\item $A: \Psi \then \Phi$, saying that $A$ knows that $\Phi$ is valid whenever the description $\Psi$ of $\psi$ is valid. 
\end{itemize}
The two formulas are semantically equivalent because the formula $\Psi$ is a complete description of the process $\psi$. The PDL assertions thus usually appear in the form $A:\Psi \then \Phi$, where $\Psi$ describes $A$'s view of the run, and $\Phi$  her conclusion about it.

The examples follow.

\subsection{Communication axioms}
The statements of PDL describe the events that happen in a run of a process, and their order. The basic PDL statements are its axioms, which we describe next. They are taken to be valid in all runs of all processes. The other valid statements are derived from them.

\subsubsection{Origination}\label{orig-ax}
The origination axioms say that any message that is received must have been sent, and that any {\sample} that is {\probe}ed must have been {\appear}ted to. This has been explained early in Sec.\ref{Processes-sec}. More precisely, any principal that controls a configuration $P$ where a message is received knows that it must have been sent by someone, no later than it was received; and similarly for a {\sample} that is {\prob}ed. Formally 
\begin{align}
\contr P  \ : \ \recv t _P  & \then \ \exists X.\ \send t_{X} \rar \recv t_P  \tagg{orig.m}\\
\contr P \ : \ \prb t _P  & \then \ \exists X.\ \apr t_{X} \rar \prb t_P  \tagg{orig.s}
\end{align}

\medskip
\subsubsection{Freshness}\label{freshness-sec}
In Sec.~\ref{MesAlg} we explained the idea of modeling random values as the indeterminates in polynomial algebras of messages. The freshness axiom extends this idea to processes, by requiring that each indeterminate $x$ must be 
\begin{itemize}
\item \emph{freshly generated} by an action $\nu[x]$ before it is used anywhere; and 
\item that it can only be used elsewhere after it has passed  in a message or a {\sample}.
\end{itemize}
which formally becomes 
\begin{alignat}{4}
\contr P &:  & a[t.x]_P  & \ \ \then\ \   \exists X.\ \ \nu[x]_{X} \hspace{6.6em} \rar\hspace{5.4em} a[t.x]_P  \tagg{fresh.1}\\
\contr P & : & \neg \nu[x]_P\ \wedge\  a[t.x]_P  &\ \ \then\ \  \exists X.\ \big(\nu[x]_{X} \rar \orig{\ssend x_{X}} \rar \rrecv x_P\rar  a[t.x]_P \big)\notag\\
&&& \hspace{4em} \vee\  \big(\nu[x]_{X} \rar \orig{\aapr x_{X}}\, \rar \pprb x_P\rar  a[t.x]_P \big) \tagg{fresh.2}  
\end{alignat}

where, using the easy subterm order $\sqsubseteq$ from  Sec.~\ref{MesAlg},
\begin{itemize}
\item  $\ssend x_X$ abbreviates $\exists t.\  x\sqsubseteq t \ \wedge\  \send t_X$, 
\item $\rrecv x_X$ abbreviates $\exists t.\  x\sqsubseteq t \ \wedge\  \recv t_X$, etc. 
\end{itemize}

\subsection{Authentication axioms}
In classical logic, a statement may be true or false. In classical formal methods, an assertion about a computation is also either true or false, and it is assumed that we can observe which one it is. The idea is that we can, e.g. inspect the program variables in the debug mode. In network computation, an assertion is still either true or false --- yet none of the participants may be able to observe whether it is true or false.  E.g., when Alice receives a message on a cyber channel, she may not be able to verify whether the statement \emph{``This message is from Bob"} is true or false. The process of verifying such non-local statements by local means is \emph{authentication}. In our model, there are two forms of authentication:
\begin{itemize}
\item interactions along authentic channels, and
\item challenge-response authentication.
\end{itemize}

\subsubsection{Interactions along authentic channels}\label{authchans-sec}
An authentic channel allows at least one of the participants to observe not only the events on their own end of the channel, but also on the other end. So there are four types of authentic channels, supporting the following assertions:

\begin{minipage}[b]{0.4\linewidth}
\centering
\begin{align}
\contr P\ :\ \send t_P  & \rar \recv t_Q  \tagg{auch.m.1}\\
\contr Q\ :\ \send t_P  & \rar \recv t_Q  \tagg{auch.m.2}
\end{align}
\end{minipage}
\hspace{.1\linewidth}
\begin{minipage}[b]{0.4\linewidth}
\centering
\begin{align}
\contr P\ :\ \apr t_P  & \rar \prb t_Q  \tagg{auch.p.1}\\
\contr Q\ :\ \apr t_P  & \rar \prb t_Q  \tagg{auch.p.2}
\end{align}
\end{minipage}

\vspace{.5\baselineskip}

Channels that satisfy \textsf{auch.m.1} or \textsf{auch.p.1} are called \emph{write}-authentic; channels that satisfy \textsf{auch.m.2} or \textsf{auch.p.2} are called \emph{read}-authentic. Here are some examples from each family:
\begin{itemize}
\item A keyboard channel guarantees to the sender that the device at which she is typing is receiving the message, and thus satisfies \textsf{(auch.m.1)}. 
\item A visual channel used for sending a message allows the receiver to see the sender, and satisfies \textsf{(auch.m.2)}.
\item When my fingerprints are taken, I observe that they are taken, and can see who is taking them, so this biometric channel satisfies \textsf{(auch.p.1)}.
\item Moreover, the person taking my fingerprints also observes that they are taking my fingerprints, so \textsf{(auch.p.2)} is also satisfied.
\item If a visual channel is used for surveying, then the surveyor sees where the display appears, and thus satisfies \textsf{(auch.p.2)} as well; etc.
\end{itemize}

Besides these assertions about the order of events, some authentic channels support other assertions. They are usually application specific, and we impose them as procedure specific axioms.

\paragraph{Authenticity of self-sampling.} One particular authenticity axiom worth mentioning is the statement that the self-observation channel is authentic, at least for sampling the sampling actions, i.e.
\begin{align}\tagg{cog}
\contr P\ :\ \  \Big(\!\!:\, \big[\prb{}_P\big]\, :\!\!\Big)_P  & \then \prb{}_P  
\end{align}
In other words, ``If I observe that I have observed something, then I have really observed something". This is the PDL version of Descartes' authentication of the world: \emph{``Cogito, ergo sum"}.


\subsubsection{Challenge-response authentication}
The challenge-response axiom is in the form
\begin{align}
\contr P\ :\ \Local_P & \then \  \Global_{PQ} \tagg{cr}
\end{align}
where, using the notation from Sec.~\ref{freshness-sec}
\bear
\Local_P\ \  & = &\ \  \nu[x]_P \rar \send{c^{PQ}x}_P \hspace{7.4em} \rar \hspace{7.4em} \recv{r^{PQ}x}_P\\
\Global_{PQ}\ \  & = &\ \   \nu[x]_P \rar \send{c^{PQ}x}_P \rar \rrecv{c^{PQ}x}_Q \rar \orig{\ssend{r^{PQ}x}_{Q}} \rar \recv{r^{PQ}x}_P
\eear
Translated into words, \textsf{(cr)} says that the owner $\contr P$ of the configuration $P$ knows that
\begin{itemize}
\item if he generates a fresh $x$, sends the challenge $c^{PQ}x$, and receives the response $r^{PQ}x$,
\item then $Q$ must have received a message containing $c^{PQ}x$ after he sent it, and then she must have sent a message containing $r^{PQ}x$ before he received it.
\end{itemize}
Using \textsf{(cr)}, from certain observations of the local events at $P$, the principal $\contr P$ can thus draw the conclusions about certain non-local events at $Q$, which he cannot directly observe. Fig.~\ref{CR-deriv} shows depicts this reasoning diagrammatically. This axiom should be viewed as an assertion about the functions $c^{PQ}$ and $r^{PQ}$. They must be such that $Q$ can compute $r^{PQ}x$ from $c^{PQ}x$, but no one else can do it.\footnote{In the cases when $c^{PQ}$ and $r^{PQ}$ are based on a secret shared between $P$ and $Q$, then $P$ can compute $r^{PQ}$ as well. In such cases, the soundness of \textsf{(cr)} depends on $\contr P$'s observation that $P$ has not done that.}

\newcommand{\someone}{Q}
\renewcommand{\deviceB}{P}
\renewcommand{\sendx}{\scriptstyle\send{c^{PQ} x}}
\newcommand{\recvresp}{\scriptstyle \recv{r^{PQ}x}}
\newcommand{\ssendresp}{\scriptstyle \ssend{r^{PQ}x}}
\renewcommand{\rrecvx}{\scriptstyle\rrecv{c^{PQ} x}}
\newcommand{\oone}{\scriptstyle 1}
\newcommand{\ttwo}{\scriptstyle 2}
\newcommand{\tthree}{\scriptstyle 3}
\newcommand{\ffour}{\scriptstyle 4}

\begin{figure}
\begin{minipage}[b]{0.4\linewidth}
\centering
\def\JPicScale{.55}
\input{PICS/CR-deriv}
\caption{The graphic view of {\sf (cr)} axiom}
\label{CR-deriv}
\end{minipage}
\hspace{1.3cm}
\begin{minipage}[b]{0.4\linewidth}
\centering
\def\JPicScale{.55}
\renewcommand{\sendx}{\scriptstyle\send{x}}
\renewcommand{\recvresp}{\scriptstyle \recv{u|V^Q.u.x}}
\renewcommand{\ssendresp}{\scriptstyle \ssend{\varsigma^{P}x}}
\renewcommand{\rrecvx}{\scriptstyle\rrecv{x}}
\renewcommand{\someone}{X=Q}
\input{PICS/S-deriv}
\caption{Challenge-response using signatures}
\end{minipage}
\end{figure}

\paragraph{Remark.} The \textsf{(cr)} axiom, and the corresponding protocol template, displayed on Fig.~9, has been one of the crucial tools of the Protocol Derivation Logic, all the way since \cite{PavlovicD:ESORICS04,PavlovicD:CSFW05}, through to \cite{PavlovicD:MFPS10}.

\section{Examples of reasoning in PDL}\label{Examples-sec}
\subsection{On the diagrammatic method}
In its diagrammatic form depicted on Fig.~9, axiom \textsf{(cr)} says that the verifier $P$, observing the local path on the left, can derive the path around the non-local actions on the right. This pattern of reasoning resembles the categorical practice of \emph{diagram chasing} \cite{MacLane,PavlovicD:MapsII}. Categorical diagrams are succinct encodings of lengthy sequences of equations. Just like the two sides of the implication in \textsf{(cr)} correspond to two paths around Fig.~9, the two sides of an equation are represented in a categorical diagram as two paths around a face of that diagram. The components of the terms in the equations correspond to the individual arrows in the paths. The equations can be formally reconstructed from the diagrams. Moreover, the diagrams can be formally combined into new proofs. The algebraic structures are thus formally transformed into geometric patterns. After some practice, the geometric intuitions begin to guide algebraic constructions in the formal language of diagrams. We apply a similar strategy to PDL.

\subsection{Cryptographic (single-factor) authentication}
We begin with a very simple example of diagrammatic reasoning, present already in \cite{PavlovicD:ESORICS04}.
\begin{thm*}
The functions
\[ c^{PQ} x = x \qquad \qquad \qquad \qquad r^{PQ} x = \varsigma^{Q}x \]
implement {\sf (cr)}, provided that the abstract signature function $\varsigma$ satisfies the following axioms:
\begin{anumerate}
\item $\varsigma^Qu = \varsigma^Qv \then u = v$, i.e., $\varsigma^Q $ is injective,
\item $\orig{\wwrit{\varsigma^{Q}t}_{X}} \then X=Q$, i.e.,  $\varsigma^{Q} t$ must originate from $Q$,
\item $V^Q.u.t\iff u = \varsigma^Q t$, i.e.,  the predicate $V^Q$ is satisfied just for the pairs $u,t$ where $u = \varsigma^Q t$,
\end{anumerate}
and that these axioms are known to the principal Bob  $ = \contr P$.
\end{thm*}

\begin{proof}
To prove the claim, we chase the diagram on Fig.~10. The numbered arrows arise from the following steps:
\begin{enumerate}
\item Bob $=\contr P$ observes $\nu[x]_{P} \rar \send x_{P} \rar \recv{r|V^Q r x}$, i.e. after sending a fresh value $x$, he receives a response $u$ which passes the verification $V^Q r x$.
\item Using the axioms (c) and \textsf{(orig.m)}, he concludes that there is some $X$ such that $\send{V^Q x}_{X} \rar \recv{r|V^Q r x}_{P}$.
\item Using \textsf{(fresh.2)} he further derives that for the same $X$ holds $\send x_{X} \rar \rrecv x_X \rar \send{V^Q x}_{X}$.
\item Using (a) and (b), Bob concludes that $V^Q x$ must have originated from $Q$. 
\end{enumerate}
Bob can, of course, only be sure that $Q$ was online between his $\send x_{P}$ and $\recv{r|V^Q r x}$, and \emph{not} that Alice = $\contr Q$ really intended to respond to his challenge. It is well known that this form of authentication is open to impersonation, since $r = V^Q x$ contains no reference to Bob or to $P$. 
\end{proof}

\subsection{Pervasive (two-factor) authentication}
Next we describe how Bob the bank authenticates Alice the customer in the CAP procedure. 

\begin{thm*}
The procedure on  Fig.~\ref{cap-run} implements authentication, i.e. satisfies \textsf{(cr)}, provided that the following assumptions are true, and known to Bob:
\begin{anumerate}
\item $Hu = Hv \then u = v$, i.e., $H$ is injective;
\item $\orig{\wwrit{s^{AB}}_{X}} \then X=S_A\vee X = C_B$, i.e.,  $s^{AB}$ must originate from $S_A$ or $C_B$;
\item $\orig{\wwrit{p^{A}}_{X}} \then X=\Human_A\vee X = S_A$, i.e.,  $p^A$ must originate from $\Human_A$ or $S_A$;
\item $\send{Hs^{AB}x}_Q\ \  \then \ \  \left(\recv{p^A,x}_Q \rar \send{Hs^{AB}x}_Q\right) \wedge p^A=\overline p^A $ , i.e.,  $S_A$ and $R$ are honest.
\end{anumerate}
\end{thm*}

\begin{proof}
Prior to the displayed execution, Alice is assumed to have sent to Bob her identity, and a request to be authenticated. Following this request, Bob's computer $C_B$ has extracted the secret $s^{AB}$ from a store, which he will use to verify that $S_A$ has generated the response.

To prove the claim, we chase the diagram on Fig.~\ref{cap-deriv}. The enumerated steps in the diagram chase correspond to the following steps in Bob's reasoning:

\renewcommand{\confA}{Q}
\renewcommand{\deviceA}{C_A}
\renewcommand{\deviceB}{C_B}
\renewcommand{\humanA}{I_A}
\renewcommand{\cardA}{S_A}
\renewcommand{\shake}{R}
\renewcommand{\vis}{\mbox{\footnotesize vis}}
\renewcommand{\bio}{\mbox{\footnotesize kyb}}
\renewcommand{\cyb}{\mbox{\footnotesize cyb}}
\renewcommand{\newx}{\scriptstyle \nu[x]}
\renewcommand{\sendx}{\scriptstyle\send x}
\renewcommand{\recvx}{\scriptstyle\recv x}
\newcommand{\sendpx}{\scriptstyle\send {p^A,x}}
\newcommand{\recvpx}{\scriptstyle\recv {p^A,x}}
\newcommand{\paover}{\scriptstyle\overline p^A}
\newcommand{\panoover}{\scriptstyle p^A}
\newcommand{\pintest}{\scriptstyle (p^A = \overline p^A)}
\newcommand{\nonce}{\scriptstyle x}
\newcommand{\sab}{\scriptstyle s^{AB}}
\newcommand{\response}{\scriptstyle \left(Hs^{AB}x\right)}
\newcommand{\sendresp}{\scriptstyle \send{Hs^{AB}x}}
\renewcommand{\rrecvx}{\scriptstyle \rrecv x}
\renewcommand{\ssendresp}{\scriptstyle \ssend{Hs^{AB}x}}
\renewcommand{\recvresp}{\scriptstyle \recv{Hs^{AB}x}}
\renewcommand{\someone}{X}
\renewcommand{\one}{\scriptstyle 1}
\newcommand{\two}{\scriptstyle 2}
\newcommand{\three}{\scriptstyle 3}
\newcommand{\four}{\scriptstyle 4}
\newcommand{\five}{\scriptstyle 5}
\newcommand{\six}{\scriptstyle 6}
\newcommand{\seven}{\scriptstyle 7}
\newcommand{\eight}{\scriptstyle 8}
\newcommand{\nine}{\scriptstyle 9}
\newcommand{\ten}{\scriptstyle 10}
\newcommand{\convresp}{\scriptstyle \apr{{Hs^{AB}x}}}
\newcommand{\sampleresp}{\scriptstyle \prb{\, Hs^{AB}x\, }}

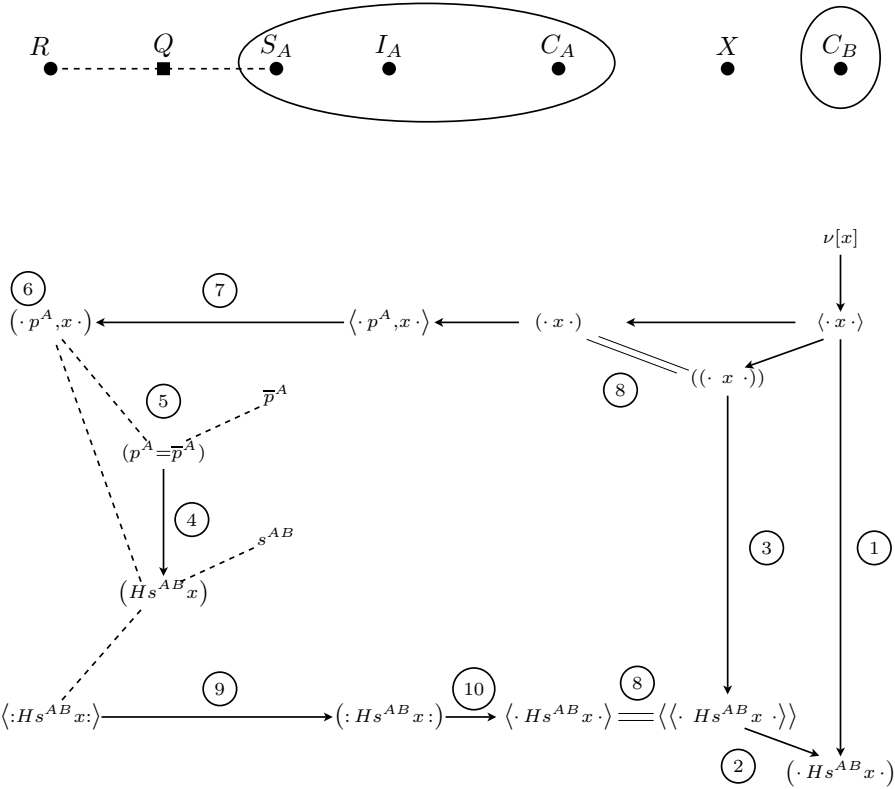
\begin{figure}
\begin{center}
\def\JPicScale{.75}
\ifx\JPicScale\undefined\def\JPicScale{1}\fi
\psset{unit=\JPicScale mm}
\psset{linewidth=0.3,dotsep=1,hatchwidth=0.3,hatchsep=1.5,shadowsize=1,dimen=middle}
\psset{dotsize=0.7 2.5,dotscale=1 1,fillcolor=black}
\psset{arrowsize=1 2,arrowlength=1,arrowinset=0.25,tbarsize=0.7 5,bracketlength=0.15,rbracketlength=0.15}
\begin{pspicture}(0,0)(157,146.63)
\rput(70,139){$\humanA$}
\rput(100,139){$\deviceA$}
\rput(8,139){$\shake$}
\rput(150,139){$\deviceB$}
\rput(30,139){$\confA$}
\psline[linestyle=dashed,dash=1 1](50,135)(10,135)
\rput{0}(150,135){\psellipse[fillstyle=solid](0,0)(1.06,-1.06)}
\rput{0}(10,135){\psellipse[fillstyle=solid](0,0)(1.06,-1.06)}
\rput{0}(100,135){\psellipse[fillstyle=solid](0,0)(1.06,-1.06)}
\rput{0}(70,135){\psellipse[fillstyle=solid](0,0)(1.06,-1.06)}
\rput{-0.12}(76.66,136.06){\psellipse[](0,0)(33.34,-10.5)}
\rput{0}(50,135){\psellipse[fillstyle=solid](0,0)(1.06,-1.06)}
\rput(50,139){$\cardA$}
\rput{90}(150,137){\psellipse[](0,0)(9,-7)}
\pspolygon[fillstyle=solid](29,136)(31,136)(31,134)(29,134)
\psline[linestyle=dashed,dash=1 1](27,69)(12,87)
\psline{->}(150,102)(150,91.88)
\rput(150,105){$\newx$}
\rput(150,90){$\sendx$}
\rput(100,90){$\recvx$}
\rput(70,90){$\sendpx$}
\psline{->}(142,90)(112,90)
\psline{->}(93,90)(78,90)
\rput(10,90){$\recvpx$}
\psline{->}(62,90)(18,90)
\rput(50,77){$\paover$}
\rput(50,52){$\sab$}
\rput(30,42){$\response$}
\rput(30,67){$\pintest$}
\rput(100,20){$\sendresp$}
\psline[linestyle=dashed,dash=1 1](26,44)(11,86)
\psline[linestyle=dashed,dash=1 1](47,75)(34,69)
\psline{->}(19,20)(60,20)
\psline{->}(30,64)(30,45)
\rput(130,20){$\ssendresp$}
\rput(130,80){$\rrecvx$}
\rput(150,10){$\recvresp$}
\psline{->}(150,87)(150,13)
\psline{->}(130,77)(130,24)
\psline{->}(147,87)(133,82)
\psline{->}(133,18)(146.25,13.12)
\rput{0}(130,135){\psellipse[fillstyle=solid](0,0)(1.06,-1.06)}
\rput(130,139){$\someone$}
\psline[linewidth=0.1](107.12,87.5)(123.12,81.5)
\psline[linewidth=0.1](105,87)(121,81)
\psline[linewidth=0.1](110.62,19.38)(116.88,19.38)
\psline[linewidth=0.1](110.62,20.62)(116.88,20.62)
\rput(0,0){\rput(156,50){\pscirclebox[]{$\one$}}}
\rput(0,0){\rput(131.88,11.25){\pscirclebox[]{$\two$}}}
\rput(0,0){\rput(136.88,50){\pscirclebox[]{$\three$}}}
\rput(0,0){\rput(35,55){\pscirclebox[]{$\four$}}}
\rput(0,0){\rput(30,76){\pscirclebox[]{$\five$}}}
\rput(0,0){\rput(6,96){\pscirclebox[]{$\six$}}}
\rput(0,0){\rput(40,95.62){\pscirclebox[]{$\seven$}}}
\rput(0,0){\rput(111,78){\pscirclebox[]{$\eight$}}}
\rput(0,0){\rput(114,26){\pscirclebox[]{$\eight$}}}
\psline[linestyle=dashed,dash=1 1](26,39)(12,23)
\rput(10,20){$\convresp$}
\rput(70,20){$\sampleresp$}
\psline{->}(80,20)(89,20)
\psline[linestyle=dashed,dash=1 1](46,50)(33,44)
\rput(0,0){\rput(40,25){\pscirclebox[]{$\nine$}}}
\rput(0,0){\rput(85,25){\pscirclebox[]{$\ten$}}}
\end{pspicture}
\caption{$B$'s reasoning in CAP}
\label{cap-deriv}
\end{center}
\end{figure}

\begin{enumerate}
\item Bob observes $\nu[x]_{C_B} \rar \send x_{C_B} \rar \recv{Hs^{AB}x}_{C_B}$.
\item Using \textsf{(orig.m)} he concludes that there is some $X$ such that $\send{Hs_Ax}_{X} \rar \recv{Hs^{AB}x}_{C_B}$.
\item Using \textsf{(fresh.2)} he further derives that for the same $X$ holds $\send x_{C_B} \rar \rrecv x_X \rar \send{Hs^{AB}x}_{X}$.
\item By (a) and (b), from the observation that he did not use $s^{AB}$, Bob concludes that $Hs^{AB}x$ must have originated in a configuration $Q$ containing $S_A$. 
\item By (c), $\ssend{p^A}_{\Human_A} \rar \rrecv{p^A}_Q \rar \left(p^A = \overline p^A\right) \rar \left(Hs^{AB}x\right)$, where the last action abbreviates $\left(r:=Hs^{AB}x\right)$, and we write out $r$ as $Hs^{AB}x$ in the rest of the diagram. (See Remark below.)
\item Since $Q$ had to also receive $x$ before computing the response in $\recv{p^A, x}_R\rar \left(Hs^{AB}x\right)$ follows by (d). So $\rrecv{p^A}_Q$ from 5 is $\recv{p^A, x}_R$.
\item By \textsf{(orig-m)}, there is $Y$ with $\send{p^A,x}_Y \rar \recv{p^A, x}_R$. By (e), $\ssend{p^A}_{\Human_A}$ from 5 must be $\send{p^A,x}_{\Human_A}$.
\item The fresh value $x$ has thus been sent to $Q$ by $\Human_A$. It follows that in 2 and 3 above must be $X = \Human_A$.
\item Since $A$ controls $S_A$ and $\Human_A$, and $S_A\in Q$ generated the response $Hs^{AB}x$, only $\Human_A$ could have sampled $Hs^{AB}x$ along the visual channel. 
\item Since $A$ controls $\Human_A$ and $C_A$, only $\Human_A$ could have sent $Hs^{AB}x$ to $C_A$ along the keyboard channel.
\end{enumerate}
These logical steps suffice to assure Bob that if he observes the local flow on the right in Fig.~\ref{cap-deriv}, then the non-local flow along the external boundary, all the way to the left side of the diagram and back, must have taken place. Comparing this diagrammatic conclusion with the pattern of \textsf{(cr)} on Fig.~9, we see that Bob has proven an instance of authentication. 

More explicitly, Bob's conclusion in Fig.~\ref{cap-deriv} is
\begin{multline*}
\send{x}_{C_B} \rar \recv{x}_{C_A} \send x_{C_A}\rar \recv x_{\Human_A} \rar \send{p^A,x}_{I_A} \rar \recv{p^A,x}_R \rar \\
\hspace{3em} 
(p^A = \overline p^A)_Q \rar\left(Hs^{AB}x\right)_Q \rar \apr{Hs^{AB}x}_R \rar 
\prb{Hs^{AB}x}_{I_A}\rar \\
  \send{Hs^{AB}x}_{\Human_A} \rar \recv{Hs^{AB}x}_{C_A} \rar \send{Hs^{AB}x}_{C_A} \rar \recv{Hs^{AB}x}_{C_B}
\end{multline*}
where we also added the trivial relay flows omitted from the figure, namely
\begin{itemize}
\item $\send x_{C_A}\flow{\rm cyb} \recv x_{\Human_A}$ on the way out, and
\item $\send{Hs^{AB}x}_{\Human_A} \flow{\rm kyb} \recv{Hs^{AB}x}_{C_A}$ on the way back.
\end{itemize}
Again, it is clear that this conclusion is clearly an instance of \textsf{(cr)} and thus realizes authentication.
\end{proof}

\paragraph{Remarks.} The reader may note that the store $r$, to which the response $Hs^{AB}x$ is assigned at run time, was present in Fig.~\ref{cap-run}, but elided in Fig.~\ref{cap-deriv}.  This is a minor quirk here, but we wanted to adher to the custom established in the informal protocol analyses, and propagated, e.g. through the strand space notation: stores and program variables are by convention avoided, and denoted by the terms assigned to them at runtime. Although such a term, strictly speaking, does not have a static value, and denoting the stores ready to receive it when it is evaluated by its unevaluated expression is abuse of notation --- it seems to be an eminently reasonable abuse, since it displays the path of the term during the run, whereas the names of all the local stores ready to receive it, only conceal its path. So the diagrammatic convention wins the day.

\paragraph{Honesty assumption.} It would be easy and natural to eliminate the assumption that $S_A$ is honest by storing $p^A$ also at $\deviceA$, and including it into the response. That would reduce the above reasoning to ``$S_A$ has been on the path of the message" and added a separate thread "$\Human_A$ has also been on the path of the message". We chose to present the above version as slightly more informative, albeit slightly weaker.

\paragraph{Does Bob need to be authenticated?} In practice, the attackers usually impersonate Bob, to steal Alice's credential and use that to impersonate her to the actual Bob, the bank. The most frequent form of that attack is, of course, phishing. Two factor authentication is devised to avoid that: here Alice does not give her credentials to Bob, but to the smart card reader. Why is that better? How does Alice know that reader $R$'s request for PIN is authentic? She \emph{sees} on the visual channel that the only card in $R$ is $S_A$, which she had put there herself. This is a simple but typical example of use of an authentic channel.

%
%
%
%
%
%

%
%
%
%
%
%
%
%
%
%
%
%
%
%
%
%
%
%
%
%
%
%

\subsection{Device pairing by handshake}
\renewcommand{\deviceB}{D_A}
\renewcommand{\deviceA}{I_A}
\renewcommand{\confA}{Q_A}
\renewcommand{\cardA}{D_B}
\renewcommand{\humanA}{I_B}
\renewcommand{\shake}{S}
\renewcommand{\vis}{\mbox{\footnotesize vis}}
\renewcommand{\shk}{\mbox{\footnotesize shk}}
\renewcommand{\newx}{\scriptstyle \nu[x]}
\renewcommand{\sendx}{\scriptstyle \apr x}
\renewcommand{\recvx}{\scriptstyle \prb x}
\renewcommand{\recvresp}{\scriptstyle x, {\rm ok}}
\renewcommand{\ssendresp}{\scriptstyle x, {\rm ok}}
\renewcommand{\rrecvx}{\scriptstyle \send{\trm{\prb {}}}}
\renewcommand{\one}{\scriptstyle \recv{\trm{\prb{}}}}
\renewcommand{\finger}{\scriptstyle \apr{f^A}}
\newcommand{\fingaf}{\scriptstyle \prb{f^A}}
\renewcommand{\bio}{\mbox{\footnotesize bio}}
\newcommand{\finesend}{\scriptstyle \send{\checkmark}}
\newcommand{\finerecv}{\scriptstyle \recv{\checkmark}}
\begin{figure}
\begin{center}
\def\JPicScale{.75}
\input{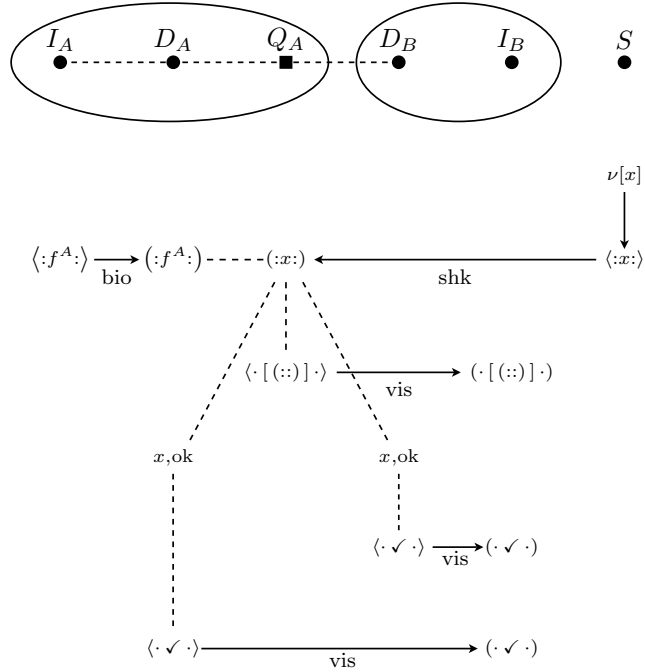}
\caption{$B$'s view of $A$'s round of device handshake}
\label{dh-deriv}
\end{center}
\end{figure}
For the final example, we return to the device handshake procedure.  For reasons of space, we omit proof of the theorem, but present that diagram that
is used to supply the proof. 

\begin{thm*}
Upon the completion of a run of the procedure described in Sec.~\ref{handshake-proc}, Bob can be sure that his and Alice's device share a key, provided that he knows that the following assumptions are valid:
\begin{anumerate}
\item $\contr P: \send{\trm{a_{Q}}}_{Q} \tto {\vis}\recv{\trm{a_{Q}}}_P$, i.e. the visual channel  satisifies {\sf (auch.m.2)} at least for events;
\item $\prb{x}_{Q_A} \then x\in \Knows_{D_A} \cap \Knows_{D_B} $, i.e., $Q_A$ distributes the same value to $D_A$ and $D_B$;
\item $\prb{t}_{Q_A} \then \nu[x]_S\rar \apr{x}_S \rar \prb{t}_{Q_A} \wedge t = x$, i.e. $Q_A$ is honest ;
\item $\send{\checkmark}_{D_A} \then {\rm ok}_{D_A} \wedge  \apr{f^A}_{I_A} \rar \prb{f^{A}}_{D_A} \rar \send{\checkmark}_{D_A} $, i.e. $D_A$ is honest; and 
\item $\send{\checkmark}_{D_B} \then  {\rm ok}_{D_B}$, i.e. $D_B$ is honest.
\end{anumerate}
\end{thm*}

Bob's formal reasoning is displayed on Fig~\ref{dh-deriv}.

\section{Conclusion, discussion, and some last minute philosophy}\label{Conclusions-sec}

\paragraph{Summary}We have presented a logical framework for reasoning about security of protocols that make use of a heterogeneous mixture of humans, devices, and channels.
We have shown how different properties of channels and configurations can be expressed and reasoned about within this framework.
A key feature of this framework is that it supports explicit reasoning about both the structure of a protocol and the contributions made by its various components,
using a combination of diagrammatic and logical methods.  Because of this, we believe that our approach can be particularly useful in giving a more rigorous foundation for white-board discussions, in which protocols are usually displayed graphically. By annotating the diagram with the proof using the methods demonstrated in this paper, formal reasoning could be brought to bear at the very earliest states of the design process.  But we also believe that it should be possible to develop tool support as well, in which a proof engine
would execute the logic, and the proof itself would be demonstrated on a graphical template.  Both avenues are a question for future research.

\paragraph{\textit{Post hoc ergo propter hoc?}} Although we speak of complete runs, as specified in Def.~\ref{run-def}, and the information flows seem completely displayed in such runs, it is always possible to raise the question whether a demonstrated temporal order of events reflects their causal order. Can it be that Bob has received the same message that Alice had sent, but that by some coincidence Carol had sent an identical message, and that Bob actually received Carol's message? If the two messages are really identical, how can we tell? 

\paragraph{Should we try to distinguish the causal connections from temporal coincidences?}  More specifically, would it be possible to refine PDL by working with statements that would specify not just the order of events, but also which interactions occurred through which channels? The idea would thus be to discern whether Bob has received Alice's or Carol's message by following the messages as they travel from channel to channel, and seeing which one goes where. 

We believe that this would just complicate logic, without bringing essentially more information. Even if we could follow two messages, with the same payload say $m_1$ and $m_2$ on their respective paths across the network, could we be sure that their paths never cross? Networks are busy places, a hop from node to node may conceal many intermediary steps, and $m_1$'s hop $a_1\to b_1$ may pass through an invisible node $x$ at the same time as $m_2$'s hop $a_2\to b_2$ passes through $x$. If that happens, then $m_1$ may emerge at $b_2$ and $m_2$ at $b_1$. Or the other way around. We will never know. It will remain uncertain which of the two messages has reached Bob in the end.

Logic has no business chasing the ghost of causality. The best we can do is describe the order of the events. Partially ordered multisets will remain a robust foundation for reasoning about network interactions and procedures for many applications to come.
%
%
%

\paragraph{Can we really model social interactions and computations within the same model?} People are not computers and society is not a computer network. How could I ever predict the behavior of a social group, when I am unable to predict the behaviors within my own family? Sometime I don't even understand my own behavior. Humans are hopelessly irrational. End of the story.

But science has strange tricks. It is unable to model the trajectory of a single particle in the air, but it models hurricanes and predicts weather. Similarly, the polling experts and the web advertisers have developed impressive techniques to predict \emph{and influence} the behaviors of large groups of people with a significant precision, although noone seems to be able to predict or influence what any particular individual will do.

Most sciences try to model reality. But people do not obey models, groups of people do not obey models, so the sciences concerned with people and with social groups largely took exception to the modeling task. It turns out, though, that the task may become easier when social networks are interleaved with networks of computers and devices, and  when people are modeled together with other social and computational actors. Can it be that models better fit people because people better fit models?

\paragraph{Acknowledgement.} The first author would like to thank Wolter Pieters \cite{PietersW:ANKH} for introducing him to Bruno Latour's ideas about actor-networks.

\bibliographystyle{plain}
\bibliography{ref-minespec,PavlovicD,ref-perdy}

\begin{thebibliography}{10}

\bibitem{Abadi-Rogaway}
M.~Abadi and P.~Rogaway.
\newblock Reconciling two views of cryptography (the computational soundness of
  formal encryption).
\newblock {\em J. of Cryptology}, 15(2):103--127, 2002.

\bibitem{Abadi-Blanchet-Lundh}
Mart\'{\i}n Abadi, Bruno Blanchet, and Hubert Comon-Lundh.
\newblock Models and proofs of protocol security: A progress report.
\newblock In Ahmed Bouajjani and Oded Maler, editors, {\em CAV}, volume 5643 of
  {\em Lecture Notes in Computer Science}, pages 35--49. Springer, 2009.

\bibitem{PavlovicD:ARSPA06}
Matthias Anlauff, Dusko Pavlovic, Richard Waldinger, and Stephen Westfold.
\newblock Proving authentication properties in the {Protocol Derivation
  Assistant}.
\newblock In Pierpaolo Degano, Ralph {K\"{u}sters}, and Luca Vigano, editors,
  {\em Proceedings of FCS-ARSPA 2006}. ACM, 2006.

\bibitem{BartheG:CSF10}
Gilles Barthe, Daniel Hedin, Santiago~Zanella B{\'e}guelin, Benjamin
  Gr{\'e}goire, and Sylvain Heraud.
\newblock A machine-checked formalization of sigma-protocols.
\newblock In {\em CSF}, pages 246--260. IEEE Computer Society, 2010.

\bibitem{Basin:decade}
David~A. Basin, Manuel Clavel, and Marina Egea.
\newblock A decade of model-driven security.
\newblock In Ruth Breu, Jason Crampton, and Jorge Lobo, editors, {\em SACMAT},
  pages 1--10. ACM, 2011.

\bibitem{Bella:book}
Giampaolo Bella.
\newblock {\em Formal Correctness of Security Protocols}.
\newblock Information Security and Cryptography. Springer Verlag, 2007.

\bibitem{BenklerY:book}
Yochai Benkler.
\newblock {\em The Wealth of Networks: How Social Production Transforms Markets
  and Freedom}.
\newblock Yale University Press, 2006.

\bibitem{Blanchet08}
Bruno Blanchet.
\newblock A computationally sound mechanized prover for security protocols.
\newblock {\em IEEE Trans. Dependable Sec. Comput.}, 5(4):193--207, 2008.

\bibitem{Blaze}
Matt Blaze.
\newblock Toward a broader view of security protocols.
\newblock In Bruce Christianson, Bruno Crispo, James~A. Malcolm, and Michael
  Roe, editors, {\em Security Protocols Workshop}, volume 3957 of {\em Lecture
  Notes in Computer Science}, pages 106--120. Springer, 2004.

\bibitem{BuhanI:safe}
Ileana Buhan, Bas Boom, Jeroen Doumen, Pieter~H. Hartel, and Raymond N.~J.
  Veldhuis.
\newblock Secure pairing with biometrics.
\newblock {\em IJSN}, 4(1/2):27--42, 2009.

\bibitem{PavlovicD:CSFW05}
Iliano Cervesato, Catherine Meadows, and Dusko Pavlovic.
\newblock An encapsulated authentication logic for reasoning about key
  distribution protocols.
\newblock In Joshua Guttman, editor, {\em Proceedings of CSFW 2005}, pages
  48--61. IEEE, 2005.

\bibitem{CortierV:algebraic}
V.~Cortier, S.~Delaune, and P.~Lafourcade.
\newblock A survey of algebraic properties used in cryptographic protocols.
\newblock {\em J. Comput. Secur.}, 14(1):1--43, 2006.

\bibitem{Cortier-Kremer:book}
V\'{e}ronique Cortier and Steve Kremer, editors.
\newblock {\em Formal Models and Techniques for Analyzing Security Protocols},
  volume~5 of {\em Cryptology and Information Security Series}.
\newblock {IOS} Press, 2011.

\bibitem{WarinschiB:survey}
V{\'e}ronique Cortier, Steve Kremer, and Bogdan Warinschi.
\newblock A survey of symbolic methods in computational analysis of
  cryptographic systems.
\newblock {\em J. Autom. Reasoning}, 46(3-4):225--259, 2011.

\bibitem{PCL}
A.~Datta, A.~Derek, J.~Mitchell, and A.~Roy.
\newblock {Protocol Composition Logic (PCL)}.
\newblock {\em Electronic Notes in Theoretical Computer Science}, 172:311--358,
  April 2007.

\bibitem{PavlovicD:MFPS03}
Anupam Datta, Ante Derek, John Mitchell, and Dusko Pavlovic.
\newblock Secure protocol composition.
\newblock {\em E. Notes in Theor. Comp. Sci.}, pages 87--114, 2003.

\bibitem{PavlovicD:JCS05}
Anupam Datta, Ante Derek, John Mitchell, and Dusko Pavlovic.
\newblock A derivation system and compositional logic for security protocols.
\newblock {\em J. of Comp. Security}, 13:423--482, 2005.

\bibitem{PavlovicD:CSFW03}
Anupam Datta, Ante Derek, John~C. Mitchell, and Dusko Pavlovic.
\newblock A derivation system for security protocols and its logical
  formalization.
\newblock In Dennis Volpano, editor, {\em Proceedings of CSFW 2003}, pages
  109--125. IEEE, 2003.

\bibitem{fuzzy-extractors}
Yevgeniy Dodis, Leonid Reyzin, and Adam Smith.
\newblock Fuzzy extractors: How to generate strong keys from biometrics and
  other noisy data.
\newblock In Christian Cachin and Jan Camenisch, editors, {\em EUROCRYPT},
  volume 3027 of {\em Lecture Notes in Computer Science}, pages 523--540.
  Springer, 2004.

\bibitem{Dolev-Yao}
Danny Dolev and Andrew~C. Yao.
\newblock On the security of public key protocols.
\newblock {\em Information Theory, IEEE Transactions on}, 29(2):198--208, 1983.

\bibitem{DorigoM:book}
Marco Dorigo and Thomas St{\"u}tzle.
\newblock {\em Ant colony optimization}.
\newblock MIT Press, 2004.

\bibitem{Dorogovtsev-Mendes}
S.~N. Dorogovtsev and J.~F.~F. Mendes.
\newblock {\em Evolution of Networks: From Biological Nets to the Internet and
  WWW}.
\newblock Oxford University Press, 2003.

\bibitem{AndersonR:CAP}
Saar Drimer, Steven~J. Murdoch, and Ross~J. Anderson.
\newblock Optimised to fail: Card readers for online banking.
\newblock In Roger Dingledine and Philippe Golle, editors, {\em Financial
  Cryptography}, volume 5628 of {\em Lecture Notes in Computer Science}, pages
  184--200. Springer, 2009.

\bibitem{PavlovicD:JCS04}
Nancy Durgin, John Mitchell, and Dusko Pavlovic.
\newblock A compositional logic for proving security properties of protocols.
\newblock {\em J. of Comp. Security}, 11(4):677--721, 2004.

\bibitem{PavlovicD:CSFW01}
Nancy Durgin, John~C. Mitchell, and Dusko Pavlovic.
\newblock A compositional logic for protocol correctness.
\newblock In Steve Schneider, editor, {\em Proceedings of CSFW 2001}, pages
  241--255. IEEE, 2001.

\bibitem{KleinbergJ:nets-book}
D.~Easley and J.~Kleinberg.
\newblock {\em {Networks, Crowds, and Markets: Reasoning About a Highly
  Connected World}}.
\newblock Cambridge University Press, 2010.

\bibitem{EllisonC:ceremonies}
Carl Ellison.
\newblock Ceremony design and analysis.
\newblock Cryptology ePrint Archive, Report 2007/399, October 2007.

\bibitem{Meadows07}
Santiago Escobar, Catherine Meadows, and Jos{\'e} Meseguer.
\newblock {Maude-NPA}: Cryptographic protocol analysis modulo equational
  properties.
\newblock In Alessandro Aldini, Gilles Barthe, and Roberto Gorrieri, editors,
  {\em FOSAD}, volume 5705 of {\em Lecture Notes in Computer Science}, pages
  1--50. Springer, 2007.

\bibitem{strands}
{Javier Thayer} Fabrega, Jonathan Herzog, and Joshua Guttman.
\newblock Strand spaces: What makes a security protocol correct?
\newblock {\em Journal of Computer Security}, 7:191--230, 1999.

\bibitem{Floyd}
Robert~W. Floyd.
\newblock Assigning meaning to programs.
\newblock In J.T. Schwartz, editor, {\em Proceedings of the Symposium on
  Applied Maths}, volume~19, pages 19--32. AMS, 1967.

\bibitem{Gratzer}
George~A. Gratzer.
\newblock {\em Universal Algebra}.
\newblock Van Nostrand Princeton, N.J.,, 1968.

\bibitem{Kozen:dynlog-book}
David Harel, Dexter Kozen, and Jerzy Tiuryn.
\newblock {\em Dynamic Logic}.
\newblock MIT Press, Cambridge, MA, 2000.

\bibitem{Hoare:ProgPred}
C.~A.~R. Hoare.
\newblock Programs are predicates.
\newblock {\em Phil. Trans. R. Soc. Lond.}, A 312:475--489, 1984.

\bibitem{Hoepman05}
Jaap-Henk Hoepman.
\newblock Ephemeral pairing on anonymous networks.
\newblock In Dieter Hutter and Markus Ullmann, editors, {\em SPC}, volume 3450
  of {\em Lecture Notes in Computer Science}, pages 101--116. Springer, 2005.

\bibitem{TygarJ:ceremonies}
Chris Karlof, J.~D. Tygar, and David Wagner.
\newblock Conditioned-safe ceremonies and a user study of an application to web
  authentication.
\newblock In {\em Proceedings of the 5th Symposium on Usable Privacy and
  Security}, SOUPS '09, pages 38:1--38:1, New York, NY, USA, 2009. ACM.

\bibitem{Meadows-Millen-Kemmerer}
Richard~A. Kemmerer, Catherine Meadows, and Jonathan~K. Millen.
\newblock Three system for cryptographic protocol analysis.
\newblock {\em J. Cryptology}, 7(2):79--130, 1994.

\bibitem{Saxena:survey}
Arun Kumar, Nitesh Saxena, Gene Tsudik, and Ersin Uzun.
\newblock A comparative study of secure device pairing methods.
\newblock {\em Pervasive Mob. Comput.}, 5:734--749, December 2009.

\bibitem{Lamport:Time}
Leslie Lamport.
\newblock Time, clocks, and the ordering of events in a distributed system.
\newblock {\em Commun. ACM}, 21(7):558--565, 1978.

\bibitem{Latour:reassembling}
Bruno Latour.
\newblock {\em Reassembling the Social: An Introduction to Actor-Network
  Theory}.
\newblock Oxford University Press, 2005.

\bibitem{Laur-Nyberg}
Sven Laur and Kaisa Nyberg.
\newblock Efficient mutual data authentication using manually authenticated
  strings.
\newblock In David Pointcheval, Yi~Mu, and Kefei Chen, editors, {\em CANS},
  volume 4301 of {\em Lecture Notes in Computer Science}, pages 90--107.
  Springer, 2006.

\bibitem{Laur-Passini}
Sven Laur and Sylvain Pasini.
\newblock User-aided data authentication.
\newblock {\em IJSN}, 4(1/2):69--86, 2009.

\bibitem{MacLane}
Saunders Mac~Lane.
\newblock {\em Categories for the Working Mathematician}.
\newblock Number~5 in Graduate Texts in Mathematics. Springer-Verlag, 1971.

\bibitem{Mayrhofer:shake}
Rene Mayrhofer and Hans Gellersen.
\newblock Shake well before use: Intuitive and secure pairing of mobile
  devices.
\newblock {\em IEEE Trans. Mob. Comput.}, 8(6):792--806, 2009.

\bibitem{McCullough:hookup}
D.~McCullough.
\newblock A hookup theorem for multilevel security.
\newblock {\em IEEE Transactions on Software Engineering}, 16(6):563--568,
  1990.

\bibitem{MeadowsC:CSFW91}
Catherine Meadows.
\newblock The {NRL Protocol Analysis Tool}.
\newblock In {\em Proceedings of the Computer Security Foundations Workshop},
  page 227. IEEE, 1991.

\bibitem{Meadows94}
Catherine Meadows.
\newblock A model of computation for the {NRL Protocol Analyzer}.
\newblock In {\em Proceedings of the Computer Security Foundations Workshop},
  pages 84--89. IEEE, 1994.

\bibitem{Meadows96}
Catherine Meadows.
\newblock The {NRL Protocol Analyzer}: {An} overview.
\newblock {\em J. Log. Program.}, 26(2):113--131, 1996.

\bibitem{PavlovicD:ESORICS04}
Catherine Meadows and Dusko Pavlovic.
\newblock Deriving, attacking and defending the {GDOI} protocol.
\newblock In Peter Ryan, Pierangela Samarati, Dieter Gollmann, and Refik Molva,
  editors, {\em Proceedings of ESORICS 2004}, volume 3193 of {\em Lecture Notes
  in Computer Science}, pages 53--72. Springer Verlag, 2004.

\bibitem{PavlovicD:dist06}
Catherine Meadows, Radha Poovendran, Dusko Pavlovic, LiWu Chang, and Paul
  Syverson.
\newblock Distance bounding protocols: authentication logic analysis and
  collusion attacks.
\newblock In R.~Poovendran, C.~Wang, and S.~Roy, editors, {\em Secure
  Localization and Time Synchronization in Wireless Ad Hoc and Sensor
  Networks}. Springer Verlag, 2006.

\bibitem{NewmanM:book}
Mark Newman.
\newblock {\em Networks: An Introduction}.
\newblock Oxford University Press, 2010.

\bibitem{Roscoe}
Long~Hoang Nguyen and Andrew~William Roscoe.
\newblock Authentication protocols based on low-bandwidth unspoofable channels:
  a comparative survey.
\newblock {\em Journal of Computer Security}, 2011.
\newblock to appear.

\bibitem{networks-biology}
Bernhard~O. Palsson.
\newblock {\em Systems Biology: Properties of Reconstructed Networks}.
\newblock Cambridge University Press, 2006.

\bibitem{Passini06:SAS}
Sylvain Pasini and Serge Vaudenay.
\newblock Sas-based authenticated key agreement.
\newblock In Moti Yung, Yevgeniy Dodis, Aggelos Kiayias, and Tal Malkin,
  editors, {\em Public Key Cryptography}, volume 3958 of {\em Lecture Notes in
  Computer Science}, pages 395--409. Springer, 2006.

\bibitem{PavlovicD:MapsII}
Dusko Pavlovic.
\newblock Maps {II}: Chasing diagrams in categorical proof theory.
\newblock {\em J. of the IGPL}, 4(2):1--36, 1996.

\bibitem{PavlovicD:SEFM10}
Dusko Pavlovic.
\newblock The unreasonable ineffectiveness of security engineering: {An
  overview}.
\newblock In Jos\'e~Luiz Fiadeiro and Stefania Gnesi, editors, {\em Proceedings
  of IEEE Conference on Software Engineering and Formal Methods, Pisa, Italy,
  2010}, pages 12--18. IEEE, 2010.

\bibitem{PavlovicD:ESORICS06}
Dusko Pavlovic and Catherine Meadows.
\newblock Deriving secrecy properties in key establishment protocols.
\newblock In Dieter Gollmann and Andrei Sabelfeld, editors, {\em Proceedings of
  ESORICS 2006}, volume 4189 of {\em Lecture Notes in Computer Science}.
  Springer Verlag, 2006.

\bibitem{PavlovicD:MFPS10}
Dusko Pavlovic and Catherine Meadows.
\newblock Bayesian authentication: {Quantifying} security of the {Hancke-Kuhn}
  protocol.
\newblock {\em E. Notes in Theor. Comp. Sci.}, 265:97 -- 122, 2010.

\bibitem{PavlovicD:ASE01}
Dusko Pavlovic and Douglas~R. Smith.
\newblock Composition and refinement of behavioral specifications.
\newblock In {\em Automated Software Engineering 2001. The Sixteenth
  International Conference on Automated Software Engineering}. IEEE, 2001.

\bibitem{PavlovicD:AMAST02}
Dusko Pavlovic and Douglas~R. Smith.
\newblock Guarded transitions in evolving specifications.
\newblock In H.~Kirchner and C.~Ringeissen, editors, {\em Proceedings of AMAST
  2002}, volume 2422 of {\em Lecture Notes in Computer Science}, pages
  411--425. Springer Verlag, 2002.

\bibitem{PeltzC:orchestrations}
Chris Peltz.
\newblock Web services orchestration and choreography.
\newblock {\em Computer}, 36:46--52, October 2003.

\bibitem{PietersW:ANKH}
Wolter {Pieters}.
\newblock Representing humans in system security models: An actor-network
  approach.
\newblock {\em Journal of Wireless Mobile Networks, Ubiquitous Computing, and
  Dependable Applications}, 2(1):75--92, 2011.

\bibitem{PrattV:Pomsets}
Vaughan Pratt.
\newblock Modelling concurrency with partial orders.
\newblock {\em Internat. J. Parallel Programming}, 15:33--71, 1987.

\bibitem{SaltzerJ:end-to-end}
J.~H. Saltzer, D.~P. Reed, and D.~D. Clark.
\newblock End-to-end arguments in system design.
\newblock {\em ACM Trans. Comput. Syst.}, 2:277--288, November 1984.

\bibitem{Stajano}
Frank Stajano, Ford-Long Wong, and Bruce Christianson.
\newblock Multichannel protocols to prevent relay attacks.
\newblock In Radu Sion, editor, {\em Financial Cryptography}, volume 6052 of
  {\em Lecture Notes in Computer Science}, pages 4--19. Springer, 2010.

\bibitem{Vaudenay:SAS}
Serge Vaudenay.
\newblock Secure communications over insecure channels based on short
  authenticated strings.
\newblock In Victor Shoup, editor, {\em CRYPTO}, volume 3621 of {\em Lecture
  Notes in Computer Science}, pages 309--326. Springer, 2005.

\bibitem{WattsD:book}
Duncan~J. Watts.
\newblock {\em Six Degrees: The Science of a Connected Age}.
\newblock W. W. Norton, New York, 2003.

\bibitem{WellmanB:compnets-socnets}
Barry Wellman, Janet Salaff, Dimitrina Dimitrova, Laura Garton, Milena Gulia,
  and Caroline Haythornthwaite.
\newblock Computer networks as social networks: Collaborative work, telework,
  and virtual community.
\newblock {\em Annual Review of Sociology}, 22(1):213--238, 1996.

\end{thebibliography}
\end{document}